\documentclass[prb,twocolumn,showpacs]{revtex4}
\usepackage{color,soul}
\usepackage{amsmath}
\usepackage{dcolumn}
\usepackage{graphicx}
\usepackage{bm}
\usepackage{amssymb}
\usepackage{subfigure}
\begin{document}

\title{Instability of the $U$(1) spin liquid with a large spinon Fermi surface in the Heisenberg-ring exchange model on the triangular lattice}
\author{Jianhua Yang and Tao Li}
\affiliation{Department of Physics, Renmin University of China, Beijing 100872, P.R.China}

\begin{abstract}
It is widely believed that the $U(1)$ spin liquid with a large spinon Fermi surface(SFS state) can be realized in the spin-$\frac{1}{2}$  $J_{1}-J_{4}$ model on the triangular lattice, when the ring exchange coupling $J_{4}$ is sufficiently strong to suppress the 120 degree magnetic ordered state. This belief is supported by many variational studies on this model and seems to be consistent with the observations on the organic spin liquid materials such as $\kappa$-(BEDT-TTF)$_{2}$Cu$_{2}$(CN)$_{3}$ and EtMe$_{3}$Sb[Pd(dmit)$_{2}$]$_{2}$, which are systems close to their Mott transition and thus have large $J_{4}$. Here we show through systematic variational search that such a state is never favored in the  $J_{1}-J_{4}$ model on the triangular lattice. Instead, a state with broken spatial symmetry is favored in the most probable parameter regime for the SFS state and has an energy much lower than that of the SFS state and other proposed variational states. More specifically, we find that for $J_{4}\ge 0.09J_{1}$, the model favors a valence bond solid state with a $4\times6$ period in its local spin correlation pattern and has a variational energy that is about $5\%$ lower than that of the SFS state. This state is separated from the $\pi$-flux state for  $J_{4}\le 0.045J_{1}$ by an intermediate symmetry breaking phase with a zigzag pattern in its local spin correlation. We find that the variational phase diagram we got is in qualitative agreement with that obtained from exact diagonalization on a $6\times6$ cluster.
\end{abstract}

\maketitle

\section{Introduction}
The search of quantum spin liquid in strongly frustrated quantum magnets has lasted for more than three decades\cite{PALee1,Balents}. A quantum spin liquid is an exotic state of matter that can host excitations with fractionalized quantum number and novel exchange statistics. Such novel excitations may be responsible for some major puzzles in strongly correlated electron systems, for example, the anomalous dynamical behaviors of some highly frustrated quantum magnets and the non-Fermi liquid behavior of the cuprate superconductors\cite{Piazza,Saya,Becca,Li1,PALee2}. These novel excitations may also be used to realize topologically protected quantum computation. However, even after the extensive efforts of the community in the last three decades, we are still not sure if such an exotic state of matter is indeed realized in any real material.    

The difficulty in truly identifing a quantum spin liquid in a real material has multiple origins. As a typical example of state of matter that is beyond the description of the Landau paradigm, a quantum spin liquid lacks the conventional local order parameter to be detected experimentally. At the same time, a quantum spin liquid usually occurs only in a very small parameter region of model Hamiltonian in which mutually frustrating exchange couplings are delicately balanced with each other. Furthermore, the inevitable existence of impurities in real materials may obscure the distinction between genuine spin liquid behavior and some glassy behavior. These difficulties are all related to the lack of physical intuition on the nature and the mechanism of emergence of the quantum spin liquid. In fact, we seldom has the physical intuition to judge if a particular kind of quantum spin liquid can be realized in a specific model. 

The $U(1)$ spin liquid with a large spinon Fermi surface(referred to as the SFS state in the following) is a clearly an exception in this regard. This state can be roughly thought of as the descendant of a metallic state after a Mott transition, in which the charge excitation has already developed a gap but the Fermi surface remains intact. Such a situation is very likely to occur if the system is in the close vicinity of the Mott transition so that the multiple spin exchange coupling is large. Indeed, in triangular lattice spin liquid materials such as $\kappa$-(BEDT-TTF)$_{2}$Cu$_{2}$(CN)$_{3}$ and EtMe$_{3}$Sb[Pd(dmit)$_{2}$]$_{2}$, people do find evidence for the existence of such a quantum spin liquid\cite{Kanoda1,Kanoda2,Zhou}. The hypothetical charge neutral spinon Fermi surface manifests itself in the metal-like behavior of the magnetic susceptibility and the specific heat at low temperature, although the system is already a charge insulator. Thermal conductivity measurements aiming to detect itinerant Fermionic spinon lead to controversial results\cite{Matsuda,Taillefer,Lisy}. Similar claims of the SFS state have also been made for other triangular magnetic systems such as 1T-TaS$_{2}$ and YbMgGaO$_{4}$ \cite{PALee3,Kanigel,Arcon,Zhang}.

Motivated by these physical expectations, a large number of theoretical efforts have been devoted to the study of the spin-$\frac{1}{2}$ Heisenberg-ring exchange model(the $J_{1}-J_{4}$ model) on the triangular lattice, in which a large ring exchange coupling $J_{4}$ is introduced to frustrate the conventional 120 degree order favored by the Heisenberg exchange coupling $J_{1}$. Variational studies find that when $J_{4}$ is strong enough, the SFS state becomes the best variational ground state of the model\cite{Motrunich,SSLee1,Grover,Xu,Liu,Lijx}. This conclusion finds some support from a DMRG simulation of the model\cite{PALee4}. However, in a more recent DMRG simulation\cite{Moore}, it is found that in the most favorable parameter regime for the SFS state there is strong evidence of spatial symmetry breaking in the ground state of the model. A systematic variational investigation of the model with potential spatial symmetry breaking allowed is thus strongly called for.   

In this work, we have performed a large scale variational optimization of the spin-$\frac{1}{2}$ $J_{1}-J_{4}$ model on the triangular lattice without assuming any symmetry a prior. To tackle such a challenging numerical problem, we have proposed several improvements on the variational optimization algorithm. We find that while the SFS state is extremely stable locally within the subspace of the Fermionic resonating valence bond(RVB) states, it is never the true variational ground state of the spin-$\frac{1}{2}$ $J_{1}-J_{4}$ model on the triangular lattice. Instead, we find that in the parameter regime which is thought to be the most favorable for the SFS state, a symmetry breaking state with a $4\times6$ periodicity in its local spin correlation pattern has an energy much lower than that of the SFS state. This state is separated from the $\pi$-flux phase favored for small $J_{4}$ by an intermediate phase with a zigzag pattern in its local spin correlation. We find that such a variational phase diagram has strong similarity with that obtained from exact diagonalization(ED) calculation on small cluster.

This paper is organized as follows. In the next section, we introduce the spin-$\frac{1}{2}$ $J_{1}-J_{4}$ model studied in this work and summarize previous theoretical results about it. We then introduce the variational wave functions we adopted in our study in Sec.III. This is followed by an introduction of the optimization algorithms we used in this work in Sec.IV.  In Section V, we present our numerical results from the variational optimization. Here we will present the full variational phase diagram of the $J_{1}-J_{4}$ model and the symmetry breaking pattern of each phase in this phase diagram. A comparison with the ED phase diagram will also be presented in this section. In the sixth section, we draw conclusion from our results and discuss their implications.

\section{The $J_{1}-J_{4}$ model on the triangular lattice}
The model we study in this work is described by the following Hamiltonian
\begin{equation}
H=J_{1}\sum_{\langle i,j \rangle}\mathbf{S}_{i} \cdot \mathbf{S}_{j}+J_{4}\sum_{[i,j,k,l]}(P_{i,j,k,l}+P^{-1}_{i,j,k,l})
\end{equation} 
in which $J_{1}$ denotes the Heisenberg exchange coupling between nearest-neighboring sites of the triangular lattice, $J_{4}$ denotes the four-spin ring exchange coupling around every elementary rhombi of the triangular lattice. In the following, we will set $J_{1}=1$ as the unit of energy. This model and its various extensions have been studied by many researchers\cite{LiMing,Motrunich,Schmidt,Grover,Xu,PALee4,Seki,Moore}. We note that the value of $J_{1}$ in our notation differs by a factor of two from that adopted in Ref.[\onlinecite{Xu,Grover}]. Now we summarize previous results about this model.

When $J_{4}=0$, the model reduces to the Heisenberg model with antiferromagnetic exchange couplings between nearest-neighboring sites of the triangular lattice. It is well known that such a model possesses a 120 degree long range order in its ground state. Within the RVB framework, it was found that such a long range ordered state can be accurately described by a Bosonic RVB ansatz\cite{ZhangQ}. Using a general theorem proved by Seiji and Sorella relating Bosoinc and Fermionic RVB state on planar graph\cite{Seiji}, this Bosonic RVB state can be connected continuously to the famous $\pi$-flux phase on the triangular lattice, which is a Fermionic RVB state describing a Dirac quantum spin liquid. The short-ranged RVB state proposed by Anderson\cite{Anderson} plays a key role in establishing such a marvelous connection. Although there is no true magnetic long range order in the $\pi$-flux phase, the local spin correlation in this state is very close to that in the 120 degree ordered ground state(see Fig.5 below). In our study, we will concentrate on the subspace of spin rotational invariant Fermionic RVB state, within which the $\pi$-flux state is the best representative of the 120 degree ordered phase. As we will see later, the $\pi$-flux phase on the triangular lattice is actually the global minimum at $J_{4}=0$ within the subspace of Fermionic RVB state.       

The ground state of the model with $J_{4} \neq 0$ is much more complicated and is still under strong debate. ED study on small clusters\cite{LiMing} find that some kind of spin liquid of unknown character may be realized for large $J_{4}$. Driven by the experimental claim of possible spin liquid behavior in triangular lattice organic salt material $\kappa$-(BEDT-TTF)$_{2}$Cu$_{2}$(CN)$_{3}$\cite{Kanoda1}, the model is revisited by Moturnich in 2005 with variational Monte Carlo method\cite{Motrunich}. It is found that a $U(1)$ spin liquid state with a large spinion Fermi surface is favored for large ring exchange coupling. Similar conclusions are also reached from other variational studies in the large $J_{4}$ regime\cite{Grover,Xu,Liu,Lijx}. The intermediate phase between the 120 degree ordered phase and the SFS state is proposed to take the form of a $Z_{2}$ spin liquid state with either an extended $s$-wave, a $d_{x^{2}-y^{2}}$-wave or a $d_{x^{2}-y^{2}}+id_{xy}$-wave spinon pairing. 
  
The SFS state is believed to host Fermionic spinon excitation around the hypothetical spinon Fermi surface. This seems to be consistent with the experimental observation of a linear-in-T specific heat and a constant magnetic susceptibility on the triangular lattice organic salt material $\kappa$-(BEDT-TTF)$_{2}$Cu$_{2}$(CN)$_{3}$, which is thought to be described by the $J_{1}-J_{4}$ model. However, gauge fluctuation beyond the mean field description is argued to generate singular correction to the specific heat\cite{Motrunich,SSLee1} of the form $C_{v}\propto T^{2/3}$, which is never observed. Driven by the tension between theories and experiments, several novel spin liquids other than the SFS phase have been proposed over the years. For example, the author of Ref.[\onlinecite{Xu}] proposed a $Z_{2}$ spin liquid with a fully gapped gauge fluctuation spectrum and a spinon dispersion with quadratic band touching(QBT) at the $\Gamma$ point. The spinon excitation above the QBT point enjoys a finite density of state but is free from singular gauge field fluctuation corrections. Such a novel state is found to have a slightly lower variational energy in the intermediate region of the ring exchange coupling than both the SFS state and the nematic spin liquid mentioned above. Another proposal is to assume spinon pairing at nonzero total momentum so that the spinon Fermi surface is only partially gapped\cite{Yao}. This proposal has no support from the calculation of the $J_{1}-J_{4}$ model.

 In Ref.[\onlinecite{Li2}], we show that the singular gauge fluctuation correction around the SFS state argued before is actually a theoretical artifact of the Gaussian approximation. When we go beyond the Gaussian level, the gauge fluctuation around the SFS state can only contribute a subleading correction to the specific heat of the form $C_{v}\propto T^{2}$, even if the SFS state is indeed the true ground state of the  $J_{1}-J_{4}$ model. The new theory also provides a unified mechanism for spin fractionalization in both 1D and 2D quantum magnets. Such a new mechanism is built on the nontrivial topological character of the Gutzwiller projected mean field state, rather than the deconfinement of slave particles.
 
The $J_{1}-J_{4}$ model has also been studied by DMRG simulations\cite{PALee4,Moore}. In an attempt to account for the possible spin liquid behavior found in 1T-TaS$_{2}$\cite{PALee3,Kanigel,Arcon}, a rather complicate triangular material argued to be described by an approximate $J_{1}-J_{4}$ model at low energy, the authors of Ref.[\onlinecite{PALee4}] revisited the $J_{1}-J_{4}$ model with DMRG. They found that a paramagnetic state without any detectable symmetry breaking pattern is realized at large value of $J_{4}$. This state possesses a spin structure factor with an approximate 2$k_{F}$ peak expected for a spin liquid with a large spinon Fermi surface. However, such a claim is challenged by a more recent DMRG simulation on the same model\cite{Moore}, in which the authors report a zigzag type symmetry breaking phase in the parameter regime thought to be the most favorable for the SFS state. A more thorough investigation is thus clearly called for to determine if the SFS state can indeed be realized in this model.

 \section{The variational wave functions}
 In this work, we describe the ground state of the $J_{1}-J_{4}$ model with the Fermionic RVB state of the form
 \begin{equation}
 |f-RVB \rangle=P_{G}\sum_{ \{i_{k} ,j_{k}\} }\prod_{k=1}^{N/2}a(i_{k},j_{k})\ P_{i_{k},j_{k}}|0 \rangle
 \end{equation}
Here
 \begin{equation}
 P_{i_{k},j_{k}}=[f^{\dagger}_{i_{k},\uparrow}f^{\dagger}_{j_{k},\downarrow}-f^{\dagger}_{i_{k},\downarrow}f^{\dagger}_{j_{k},\uparrow}]
 \end{equation}
 creates a Fermionic spin singlet pair(valence bond) between site $i_{k}$ and $j_{k}$. $f^{\dagger}_{i,\sigma}$ is the Fermion creation operator on site $i$ and with spin $\sigma$. $|0 \rangle$ is the vacuum of the $f$-Fermion. $a(i_{k},j_{k})$ is the RVB amplitude of the $k$-th valence bond and satisfies $a(i_{k},j_{k})=a(j_{k},i_{k})$. $\sum_{ \{i_{k} ,j_{k}\} }$ denotes the sum over all valence bond configurations on the lattice. $P_{G}$ denotes the Gutzwiller projection introduced to enforce the single occupancy constraint on the $f$-Fermions. 
 
 \subsection{General Fermionic RVB states}
The Fermionic RVB state can be expanded in the Ising basis as
 \begin{equation}
 |f-RVB \rangle=\sum_{\{\sigma_{1},.....\sigma_{N}\}} \Psi(\sigma_{1},....,\sigma_{N})|\sigma_{1},....,\sigma_{N}\rangle
 \end{equation} 
 in which 
\begin{equation}
 |\sigma_{1},....,\sigma_{N}\rangle=\prod_{k=1}^{N/2}f^{\dagger}_{i_{k},\uparrow}f^{\dagger}_{j_{k},\downarrow}|0\rangle
 \end{equation}
 denotes an Ising basis written in terms of the Fermion Fock state
 \begin{equation}
 \Psi(\sigma_{1},....,\sigma_{N})=Det[\mathbf{ A} ]
 \end{equation}
is the corresponding wave function amplitude. Here and in the following, we will use $i_{k}$ and $j_{k}$ to denote the locations of the $k$-th up and the $k$-th down spin in the Ising basis $ |\sigma_{1},....,\sigma_{N}\rangle$, rather than the two end points of the $k$-th valence bond. $\mathbf{A}$ is a $\frac{N}{2} \times \frac{N}{2}$ matrix with its matrix element given by $a(i_{k},j_{k'})$ at its $k$-th row and $k'$-th column.

In practice, we can treat the RVB amplitude $a(i_{k},j_{k})$ as variational parameter directly. The variational state constructed in this way will be referred to as the general RVB state(gRVB) in the following discussion. The number of variational parameters in the gRVB state increases very rapidly with the system size. As we will see in the next section, such an unfavorable feature of the gRVB state is compensated partly by the fact that the calculation of the energy gradient in the gRVB state is rather cheap.

 \subsection{Fermionic RVB states generated from mean field ansatzs}
The Fermionic RVB state can also be generated by Gutzwiller projection of mean field ground state of the following Bogoliubov-de Gennes Hamiltonian
\begin{equation}
H_{MF}=-\sum_{\{i,j\},\sigma}(\chi_{i,j} f^{\dagger}_{i,\sigma}f_{j,\sigma}+h.c.)+\sum_{\{i,j\}}(\Delta_{i,j}f^{\dagger}_{i,\uparrow}f^{\dagger}_{j,\downarrow}+h.c.)
\end{equation}
Here the condition $\Delta_{i,j}=\Delta_{j,i}$ is imposed to enforce spin rotational symmetry of the variational ground state. In general, both $\chi_{i,j} $ and $\Delta_{i,j}$ can be chosen complex. In our work, $\chi_{i,j} $ and $\Delta_{i,j}$ will be chosen real and be restricted to the nearest neighboring bonds.They are otherwise free of any assumption. 

$H_{MF}$ is usually referred to as a mean field ansatz or a variational ansatz of the Fermionic RVB state. To generate $ |f-RVB \rangle$, we rewrite $H_{MF}$ in the following form
\begin{equation}
H_{MF}=\psi^{\dagger}\left(\begin{array}{cc}-\bm{\chi} & \bm{\Delta} \\\bm{\Delta}^{\dagger} & \bm{\chi}^{*}\end{array}\right)\psi=\psi^{\dagger}\mathbf{M}\psi
\end{equation}
in which 
\begin{equation}
\psi^{\dagger}=(f^{\dagger}_{1,\uparrow},...,f^{\dagger}_{N,\uparrow},f_{1,\downarrow},....,f_{N,\downarrow})
\end{equation}
Here $\bm{\chi}$ and $\bm{\Delta}$ are $N\times N$ matrices with $\chi_{i,j} $ and $\Delta_{i,j}$ as their matrix elements. $H_{MF}$ can be diagonalized by the following unitary transformation
\begin{equation}
\psi=\left(\begin{array}{cc}\bm{u} & \bm{v} \\-\bm{v} & \bm{u}\end{array}\right)\gamma
\end{equation}
in which $\bm{u}$ and $\bm{v}$ are $N\times N$ matrices. The diagonalized Hamiltonian takes the form of
\begin{equation}
H_{MF}=\gamma^{\dagger}\left(\begin{array}{cc}\bm{E} & 0 \\0 & -\bm{E}\end{array}\right)\gamma
\end{equation}
in which $\bm{E}$ is a $N\times N$ diagonal matrix with positive definite diagonal matrix elements.
When $\bm{\Delta}\neq 0$, the RVB amplitude $a(i_{k},j_{k})$ of the corresponding Fermionic RVB state is given by
\begin{equation}
a(i_{k},j_{k})=(\bm{u}^{-1}\bm{v})_{i_{k},j_{k}}.
\end{equation}
When $\bm{\Delta}=0$, the RVB amplitude is given by
\begin{equation}
a(i_{k},j_{k})=\sum_{E_{m}< \mu}\varphi^{*}_{m}(i_{k})\varphi_{m}(j_{k})
\end{equation}
in which $\varphi_{m}(j_{k})$ is the eigenvector of the matrix $\bm{\chi}$ with eigenvalue $E_{m}$. Here $\mu$ denotes the chemical potential of the Fermionic spinon at half filling.

\subsection{Particle-hole transformation}
In practical variational calculation, it is often convenient to adopt a particle-hole transformation on the down-spin Fermions, which is given by
\begin{equation}
f^{\dagger}_{i,\downarrow}\rightarrow\tilde{f}_{i,\downarrow}
\end{equation}
The mean field ground state of $H_{MF}$ is then constructed by filling up the lowest $N$ eigenvectors of $\mathbf{M}$. More specifically, we can expand the RVB state as
\begin{equation}
|f-RVB\rangle=\sum_{\{i_{1},...i_{N/2}\}}\tilde{\Psi}(i_{1},...,i_{N/2})\prod_{k=1}^{N/2}f^{\dagger}_{i_{k},\uparrow}\tilde{f}^{\dagger}_{i_{k},\downarrow}|\tilde{0}\rangle
\end{equation}
in which 
\begin{equation}
|\tilde{0}\rangle=\prod_{i=1}^{N}f^{\dagger}_{i_{k},\downarrow}|0\rangle
\end{equation}
is the reference state with all sites occupied by down spin Fermions(or the vacuum of $\tilde{f}$ operators). Here $i_{1},...,i_{N/2}$ denote the locations of the $N/2$ up spin Fermions. Note that they are also the locations of the $N/2$ holes of the down spin Fermion. 
The wave function in the particle-hole transformed picture then takes the form of
\begin{equation}
\tilde{\Psi}(i_{1},...,i_{N/2})=Det[\bm{\Phi}]
\end{equation}
in which $\bm{\Phi}$ is a $N\times N$ matrix of the form
\begin{eqnarray}
\bm{\Phi}=\left(\begin{array}{cccc}\phi_{1}(i_{1}) & . & . & \phi_{N}(i_{1}) \\. & . & . & . \\ \phi_{1}(i_{N/2}) & . & . & \phi_{N}(i_{N/2}) \\   \phi_{1}(i_{1}+N) & . & . & \phi_{N}(i_{1}+N) \\. & . & . & .\\
 \phi_{1}(i_{N/2}+N) & . & . & \phi_{N}(i_{N/2}+N)\nonumber
 \end{array}\right)
\end{eqnarray}
 Here $\phi_{n}(i)$ denotes the $i$-th component of the $n$-th eigenvector of the matrix $\mathbf{M}$ with negative eigenvalue.
 
 \subsection{Number of variational parameters}
 In this study, we use either the general Fermionic RVB state or RVB state generated from mean field ansatz to describe the ground state of the $J_{1}-J_{4}$ model. For the general RVB state, there will be $\frac{N(N-1)}{2}$ variational parameters to be optimized on a finite cluster with $N$ sites. For RVB state generated from mean field ansatz, we can choose either a $U(1)$ ansatz, in which $\bm{\Delta}=0$, or a $Z_{2}$ ansatz, in which both $\bm{\chi}$ and $\bm{\Delta}$ are nonzero. In the $U(1)$ ansatz, there are $3N$ variational parameters to be optimized, which are the hopping amplitude $\chi_{i,j}$ on the $3N$ nearest neighboring bonds. In the $Z_{2}$ ansatz, there are $6N+1$ variational parameters to be optimized, which are the hopping amplitude $\chi_{i,j}$ and the pairing amplitude $\Delta_{i,j}$ on the $3N$ nearest neighboring bonds and the chemical potential setting the Fermi level of the spinon. The gRVB state represents the most general form of a Fermionic RVB state and it may not be generated by any short-ranged mean field ansatz. Correspondingly, it contains a much larger number of variational parameters. The optimization of large number of variational parameters calls for very efficient optimization algorithm, which we will now introduce.

\section{Some new developments of variational optimization algorithm}
The key step in the variational Monte Carlo optimization is the computation of the variational energy and its gradient. Suppose that the wave function of the variational state $|\Psi\rangle$ in a orthonormal basis $|R\rangle$ is given by $\Psi(R)$, the variational energy is then given by
\begin{equation}
E=\langle H \rangle_{\Psi}=\frac{\langle\Psi| H |\Psi\rangle}{\langle \Psi |\Psi \rangle}=\frac{\sum_{R}|\Psi(R)|^2 E_{loc}(R)}{\sum_{R}|\Psi(R)|^2}
\end{equation}
in which the local energy $E_{loc}(R)$ is defined by
\begin{equation}
E_{loc}(R)=\sum_{R'}\langle R |H| R' \rangle \frac{\Psi(R')}{\Psi(R)}
\end{equation}
The gradient of the variational energy with respect to the variational parameters is given by
\begin{equation}
\nabla E= \langle \nabla \ln \Psi(R) E_{loc}(R) \rangle_{\Psi}-E\langle \nabla \ln \Psi(R) \rangle_{\Psi}
\end{equation}
Here we denotes the variational parameters as $\bm{\alpha}$ and abbreviate $\nabla_{\bm{\alpha}}$ as $\nabla$.
These two quantities can be computed by standard Monte Carlo sampling on the distribution generated by $|\Psi(R)|^2$. In the calculation of the gradient, the key quantity is $\nabla \ln \Psi(R)$. For the general RVB state, it is given by
\begin{equation}
\nabla\ln \Psi(R)=\mathbf{Tr} [\nabla \mathbf{A} \mathbf{A}^{-1}]
\end{equation}
Since we take the RVB amplitude $a(i_{k},j_{k})$ directly as the variational parameters, the matrix elements of $\nabla \mathbf{A}$ is either 1 or 0, depending on whether the gradient is taken on a given RVB amplitude. For RVB state generated from Gutzwiller projection of the ground state of a mean field ansatz, we have 
\begin{equation}
\nabla\ln \Psi(R)=\mathbf{Tr} [\nabla \bm{\Phi} \bm{\Phi}^{-1}]
\end{equation}
The matrix elements of $\nabla \bm{\Phi}$ can be calculated from the first order perturbation theory as follows
\begin{equation}
\nabla \phi_{n}=\sum_{E_{m}>0}\frac{\langle\phi_{m}|\nabla H_{MF}|\phi_{n}\rangle}{E_{n}-E_{m}}\phi_{m}
\end{equation}  
Here $|\phi_{n} \rangle$ and $E_{n}$ denote the $n$-th eigenvector and eigenvalue of the mean field Hamiltonian $H_{MF}$.

To proceed the optimization procedure, one need also the Hessian matrix of $E$ with respect to the variational parameters. However, the Hessian matrix is usually too expensive to be calculated numerically. Different variational optimization algorithms differ in their way to approximate the Hessian matrix, which we will now review briefly.

\subsection{The steepest descent}
The steepest descent(SD) algorithm is the simplest optimization algorithm. It corresponds to setting the Hessian matrix proportional to the identity matrix. In the SD algorithm, the variational parameters are updated as follows
\begin{equation}
\bm{\alpha}\rightarrow \bm{\alpha}-\delta \nabla E
\end{equation} 
 in which $\delta$ is the step length. The step length is usually adjusted by trial and error. A more intelligent way to adjust the step length is the following self-learning trick, in which the step length is updated according to the change in the direction of the gradient as follows
 \begin{equation}
 \delta\rightarrow \delta\times(1+\eta \frac{\nabla E\cdot \nabla E'}{|\nabla E||\nabla E'|})
 \end{equation}  
 Here $\eta\in (0,1)$ is an acceleration factor, $\nabla E$ and $\nabla E'$ are the gradients of the energy in the current and the previous step. Suitable choice of $\eta$ can accelerate significantly the optimization procedure at the initial stage, when the energy gradient is large.  
  
 The SD algorithm usually works very well at the initial stage of the optimization procedure. However, it loses efficiency when the optimization procedure encounters long and narrow valley with flat bottom in the energy landscape. In such a situation, there will be large fluctuation in the eigenvalues of the Hessian matrix. Approximating the Hessian matrix with an identity matrix is then clearly not a wise choice.
 
 \subsection{Stochastic reconfiguration}
 The stochastic reconfiguration algorithm is a widely used variational optimization algorithm. It mimics the effect of the Hessian matrix with a positive-definite Hermitian matrix  $\mathbf{S}$ generated from the metric of the variational state in the variational space\cite{Seiji}. More specifically, $\mathbf{S}$ is given by
 \begin{equation}
 \mathbf{S}=\langle \nabla ln\Psi(R) \nabla ln\Psi(R) \rangle_{\Psi}-\langle \nabla ln\Psi(R) \rangle_{\Psi}\langle \nabla ln\Psi(R) \rangle_{\Psi}
 \end{equation}
 It is easy to show that $\mathbf{S}$ is just the Hessian matrix of the following quantity with respect to the change of variational parameters
 \begin{equation}
 \Delta^{\mathrm{SR}}=2-2\frac{\langle\Psi|\Psi'\rangle}{\sqrt{\langle\Psi|\Psi\rangle\langle\Psi'|\Psi'\rangle}}
 \end{equation} 
  Here we assume that $|\Psi\rangle$ is fixed and $|\Psi'\rangle$ is varying. $\Delta^{\mathrm{SR}}$ defined in this way can be interpreted as the distance between $|\Psi\rangle$ and $|\Psi'\rangle$
in the Hilbert space.

 In the SR algorithm, the variational parameters are updated as follows
  \begin{equation}
\bm{\alpha}\rightarrow \bm{\alpha}-\delta\ \mathbf{S}^{-1}\nabla E
\end{equation} 
in which $\delta$ is the step length. The self-learning acceleration trick is also applicable in the SR method.  

The introduction of the $\mathbf{S}$ matrix in the SR method amounts to replace the naive distance in the Euclidean space of variational parameters with the distance in the Hilbert space. Such a regulation procedure can be very helpful when some variational parameters are nearly redundant. However, since $\mathbf{S}$ only depends on the variational state but not on the Hamiltonian, it can not approximate the effect of the true Hessian matrix correctly in certain situations. In practice, the SR method may still suffer slow convergence or even run away from true minimum. A better approximation of the effect of the Hessian matrix is needed.

\subsection{BFGS}
The Broyden-Fletcher-Goldfarb-Shanno (BFGS) method is a quasi-Newton method. It generates an iterative approximation for the (inverse) of the Hessian matrix\cite{Nocedal} from the gradient. The approximate (inverse)Hessian matrix is updated as follows
\begin{equation}
\mathbf{B}_{k+1}=\left(\mathbf{I}-\frac{\mathbf{s}_{k}\mathbf{y}^{T}_{k}}{\mathbf{y}^{T}_{k}\mathbf{s}_{k} } \right)\mathbf{B}_{k}\left(\mathbf{I}-\frac{\mathbf{y}_{k}\mathbf{s}^{T}_{k}}{\mathbf{y}^{T}_{k}\mathbf{s}_{k} } \right)+\frac{\mathbf{s}_{k}\mathbf{s}^{T}_{k}}{\mathbf{y}^{T}_{k}\mathbf{s}_{k} }
\end{equation}
in which $k=0,1,2.....$
\begin{eqnarray}
\mathbf{s}_{k}&=&\bm{\alpha}_{k+1}-\bm{\alpha}_{k}\nonumber\\
\mathbf{y}_{k}&=&\nabla E_{k+1}-\nabla E_{k}
\end{eqnarray}  
are the difference between successive variational parameters and energy gradient. $\bm{\alpha}_{k=0}$ is the initial guess of the variational parameters and $\nabla E_{k=0}$ is the energy gradient at the starting point. The Hessian matrix is initially set to be the identity matrix, namely $\mathbf{B}_{k=0}=\mathbf{I}$. 

Using such an iterative approximation on the Hessian matrix, the variational parameters are updated as follows
\begin{equation}
\bm{\alpha}_{k+1}=\bm{\alpha}_{k}+\delta\ \mathbf{B}_{k} \nabla E_{k}
\end{equation}  
Here $\delta$ is the step length. In principle the step length should be determined by a linear search in the direction of $ \mathbf{B}_{k} \nabla E_{k}$. Such a linear search can be accomplished in principle in the variational Monte Carlo simulation by reweighting in the searching direction. However, to reduce computational expense, we choose a fixed step length by trial and error.
   
\subsection{Conjugate gradient}
Another simpler method to go beyond the steepest descent method is the conjugate gradient(CG) method\cite{Nocedal}. It corrects the searching direction as follows
\begin{equation}
\mathbf{d}_{k+1}=-\nabla E_{k+1}+\beta_{k+1} \mathbf{d}_{k}
\end{equation} 
in which $k=0,1,2,....$
\begin{equation}
\beta_{k+1} =\frac{\nabla E_{k+1}\cdot \nabla E_{k+1}}{\nabla E_{k}\cdot \nabla E_{k}}
\end{equation}
or
\begin{equation}
\beta_{k+1} =max(\frac{\nabla E_{k+1}\cdot (\nabla E_{k+1}-\nabla E_{k})}{\nabla E_{k}\cdot \nabla E_{k}},0)
\end{equation}
Initially we set $\beta_{0}=0$.

The variational parameters are updated in the conjugate gradient method as follows
\begin{equation}
\bm{\alpha}_{k+1}=\bm{\alpha}_{k}+\delta\  \mathbf{d}_{k}
\end{equation}
in which the step length $\delta$ should be determined by linear search in the direction of $ \mathbf{d}_{k}$. In practice, to reduce computational expense we choose a fixed step length by trial and error.

\subsection{Comparison of the performance of different optimization algorithms}
We note that for both the BFGS and the CG method, the initial step of the optimization is just the SD update. As a result of the finite accuracy in the computation of the energy and the gradient in variational Monte Carlo simulation, we cut off the BFGS and the CG iteration at a finite depth. We find empirically that 10 steps of BFGS or CG iteration has the best balance between numerical efficiency and numerical stability, after which we restart the iteration by setting $k=0$. 

Fig.1 compares the performance of the four algorithms in a typical situation. Here we optimize the variational energy of the $J_{1}-J_{4}$ model at $J_{4}=0.065$ with the $U(1)$ mean field ansatz. We start from the same initial guess and use the same normalized step length. We find that the BFGS method has the best numerical efficiency and stability among the four algorithms. The CG algorithm exhibits a similar numerical efficiency as the BFGS algorithm but with a significantly larger fluctuation. Both the steepest descent and the stochastic reconfiguration method fail to escape from the the local minimum around $E\approx -0.4976$  within 2000 optimization steps. In the following, we will mainly use the BFGS method. However, we note that the conjugate gradient method has the advantage that it does not need to store the approximate Hessian matrix, which is huge when the number of variational parameters is large.

\begin{figure}
\includegraphics[width=8.5cm]{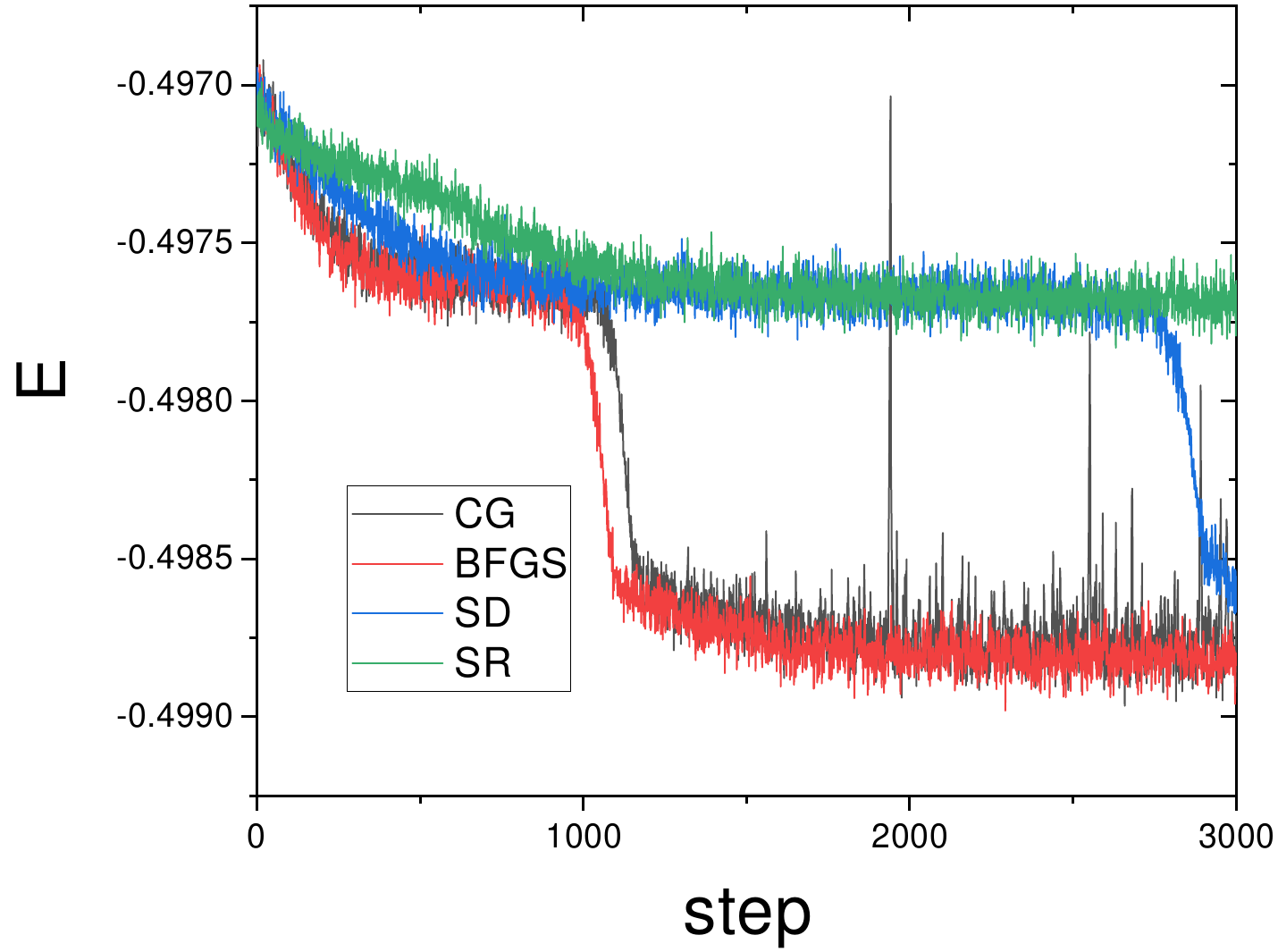}
\caption{The comparison of the performance of the four optimization algorithms in a typical situation. Shown here is the variational energy of the $J_{1}-J_{4}$ model at $J_{4}=0.065$ with a $U(1)$ mean field ansatz,  starting from the same initial guess and using the same normalized step length.}
\end{figure}

\section{The variational phase diagram of the $J_{1}-J_{4}$ model}
We have performed variational optimization of the $J_{1}-J_{4}$ model on a $L\times L$ cluster with periodic boundary condition in both the $\mathbf{a}_{1}$ and the $\mathbf{a}_{2}$ direction(see Fig.2). We have adopted both the general RVB state and the RVB state generated from a mean field ansatz of either the $U(1)$ or the $Z_{2}$ form. No further assumption is made on the form of the RVB state. Fig.3 shows the variational phase diagram we get from the optimization. For clearness we only present the result for the $Z_{2}$ RVB state. The results for the gRVB state and the $U(1)$ RVB state are qualitatively similar, with the variational energies satisfying $E_{gRVB}\le E_{Z_{2}} \le E_{U(1)}$.  

We find that there are three phases in the phase diagram of the $J_{1}-J_{4}$ model. For $0\le J_{4}\le 0.045$, the optimized RVB state is the well known $\pi$-flux phase on the triangular lattice\cite{Seiji}. The optimized variational energy exhibits only a tiny curvature in this regime, indicating that the $\pi$-flux phase, the best representative of the 120 degree ordered phase in the space of Fermionic RVB state, enjoys a finite range of stability when we increase the ring exchange coupling. For $J_{4}\ge 0.09$, the variational energy also has a very small curvature. We find that the optimized RVB state in this regime breaks the translational symmetry and exhibits a $4\times6$ periodicity in its local spin correlation pattern. This phase will thus be called the $4\times6$ phase in the following.  For $0.045\le J_{4}\le 0.09$, we find another symmetry breaking phase with zigzag spin correlation pattern. This intermediate phase will be called the zigzag phase in the following. Notably, we find that the optimized variational energy is significantly lower than that of the SFS state. In fact, the SFS state is found to be never a good approximation of the ground state of the $J_{1}-J_{4}$ model. This is the most important finding of this work.

\begin{figure}
\includegraphics[width=8.5cm]{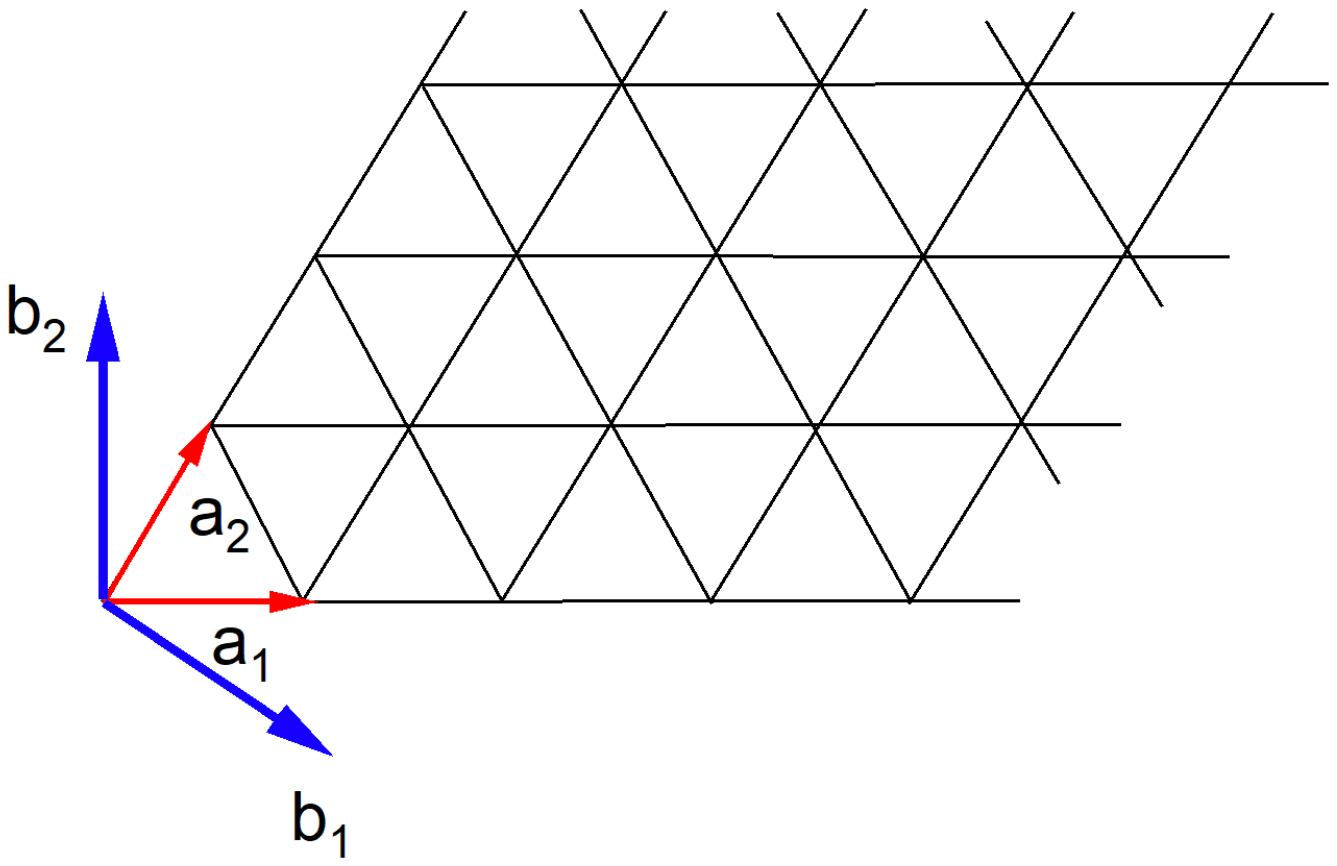}
\caption{The triangular lattice and its reciprocal vectors. Our calculation is done on a $L\times L$ cluster with periodic boundary condition in both the $\mathbf{a}_{1}$ and the $\mathbf{a}_{2}$ direction.}
\end{figure}

\begin{figure}
\includegraphics[width=8.5cm]{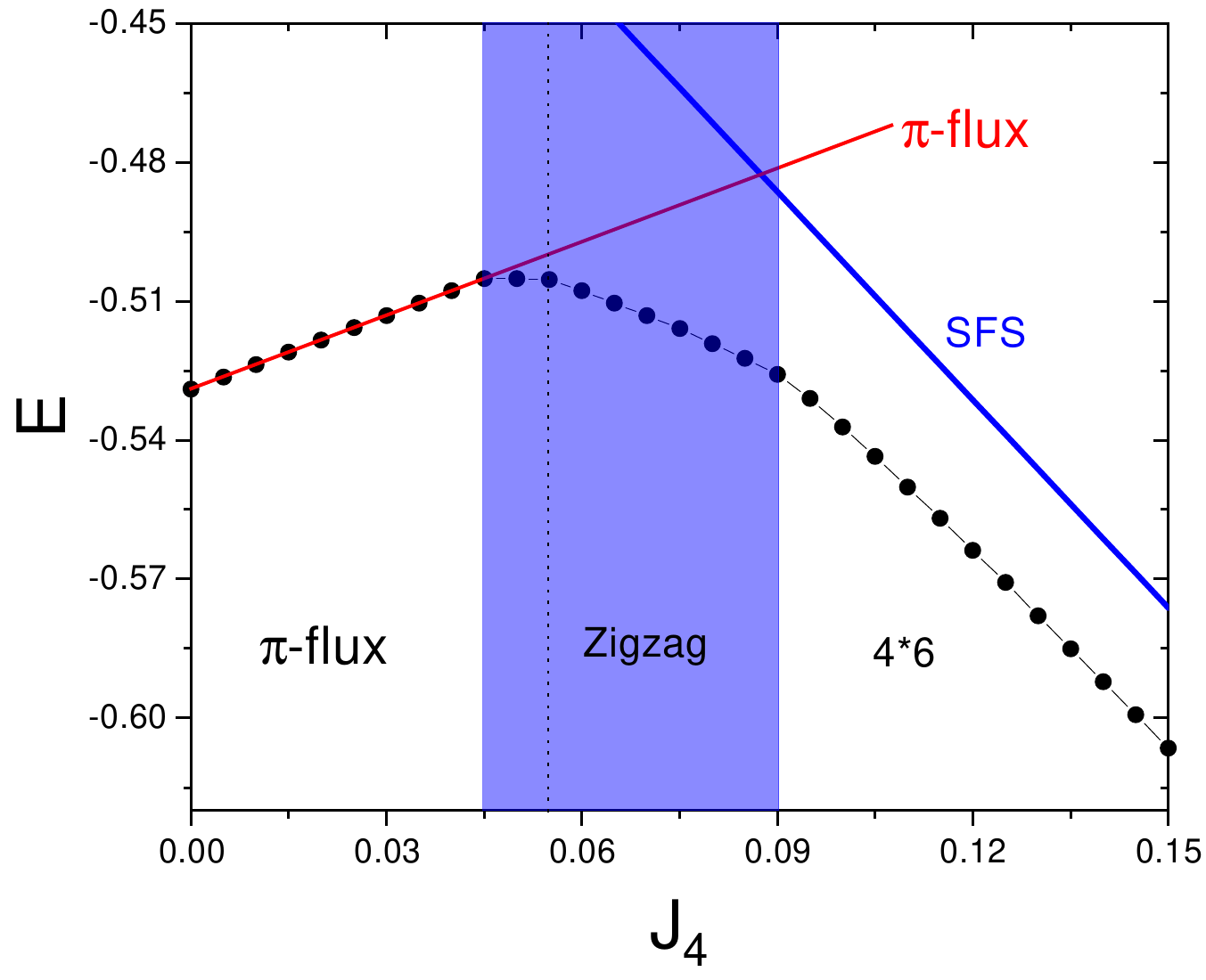}
\caption{The variational phase diagram of the $J_{1}-J_{4}$ model on the triangular lattice. Shown here is the result obtained from the $Z_{2}$ RVB state. The computation is done on a $12\times12$ cluster with periodic boundary condition. The results for the gRVB state and the $U(1)$ RVB state are qualitatively similar. Here the thick blue line and thick red line represent the variational energies of the SFS state and the $\pi$-flux phase. The $4\times 6$ phase in the large $J_{4}$ region exhibits $4\times 6$ periodicity in its local spin correlation pattern. The dashed line within the zigzag phase separates two phases with different translational symmetries.}
\end{figure}  

Below we will present the results for each of the three phases in more detail.

\begin{figure}
\includegraphics[width=8.5cm]{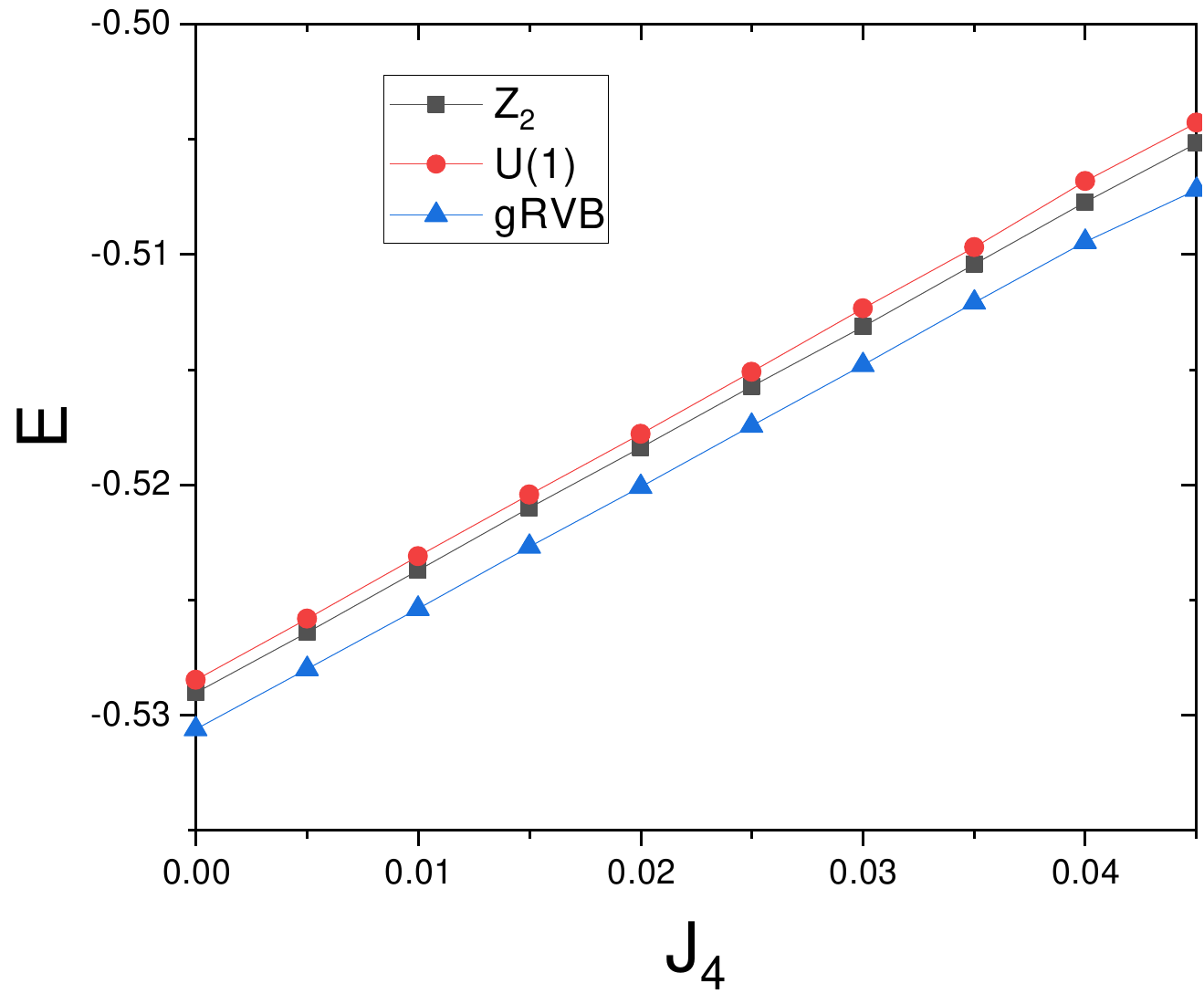}
\caption{The variational energies of the three kinds of RVB state as functions of $J_{4}$ in the range $0\le J_{4}\le 0.045$, computed on a $12\times 12$ cluster with periodic boundary condition. The variational state is essentially independent of $J_{4}$ in this regime and corresponds to the $\pi$-flux phase on the triangular lattice.}
\end{figure}  

\subsection{The $\pi$-flux phase}
Fig.4 plots the variational energies computed from the three kinds of RVB state as a function of $J_{4}$ for $0\le J_{4}\le 0.045$. The variational energies of all these three kinds of RVB state are very close to each other and are within statistical error perfect linear functions of $J_{4}$. Such a nearly zero curvature behavior in the variational energy indicates that the variational ground state is almost independent of $J_{4}$. In the case of the $U(1)$ RVB state, the optimized variational state reduces to the well known $\pi$-flux phase on the triangular lattice. This can be explicitly seen from the optimized variational parameters of the $U(1)$ ansatz, which indeed encloses a gauge flux of $\pi$ around every elementary rhombi of the triangular lattice. We find that the small difference in the optimized variational energies between the $Z_{2}$ and the $U(1)$ RVB state can be attributed to finite size effect. On a $24 \times 24$ cluster, the optimized variational energy of the two states become indistinguishable within the statistical error of the variational Monte Carlo simulation. The equivalence between the optimized $U(1)$ and $Z_{2}$
RVB state in this low $J_{4}$ regime can also be seen from the static spin structure factor $S(\mathbf{q})$, which is defined as
\begin{equation}
S(\mathbf{q})=\frac{1}{N}\sum_{i,j}\mathbf{S}_{i}\cdot\mathbf{S}_{j}\ e^{i\mathbf{q}\cdot(\mathbf{R}_{i}-\mathbf{R}_{j})}
\end{equation}
Here we define the components of the wave vector as follows
\begin{equation}
\mathbf{q}=q_{1}\mathbf{b}_{1}+q_{2}\mathbf{q}_{2}
\end{equation}
in which $\mathbf{b}_{1,2}$ are the two reciprocal vectors of the triangular lattice.
 In Fig.5, we present the static spin structure factor of the optimized $U(1)$ and $Z_{2}$ RVB state at $J_{4}=0$. We find that  $S(\mathbf{q})$ is within statistical error independent of $J_{4}$ in the whole $0\le J_{4} \le 0.045$ regime and features pronounced peaks at the wave vector corresponding to the 120 degree ordered phase, namely $\mathbf{q}=(\frac{2\pi}{3},\frac{4\pi}{3})$ and $\mathbf{q}=(\frac{4\pi}{3},\frac{2\pi}{3})$.

\begin{figure}
\includegraphics[width=8cm]{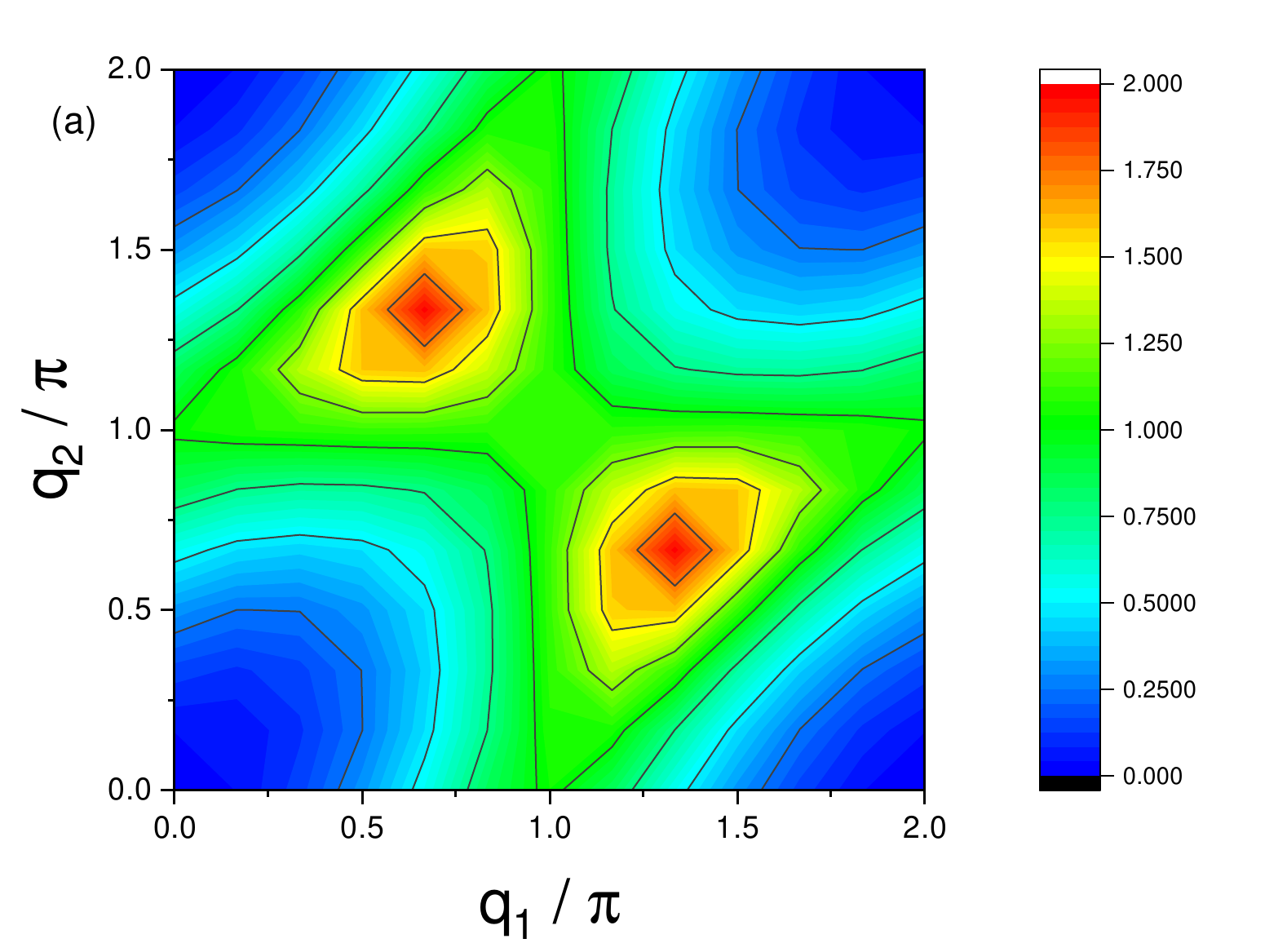}
\includegraphics[width=8cm]{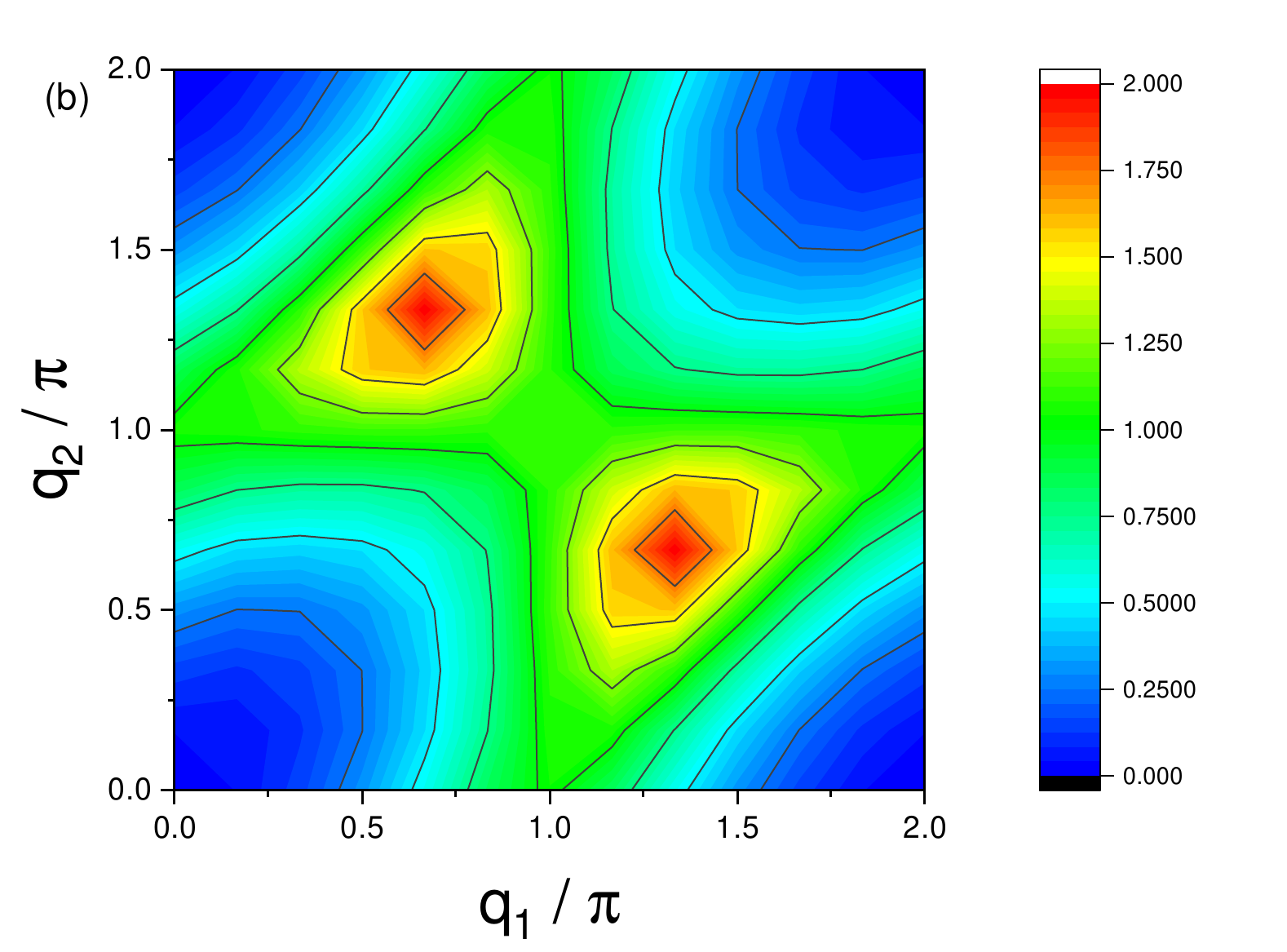}
\includegraphics[width=8cm]{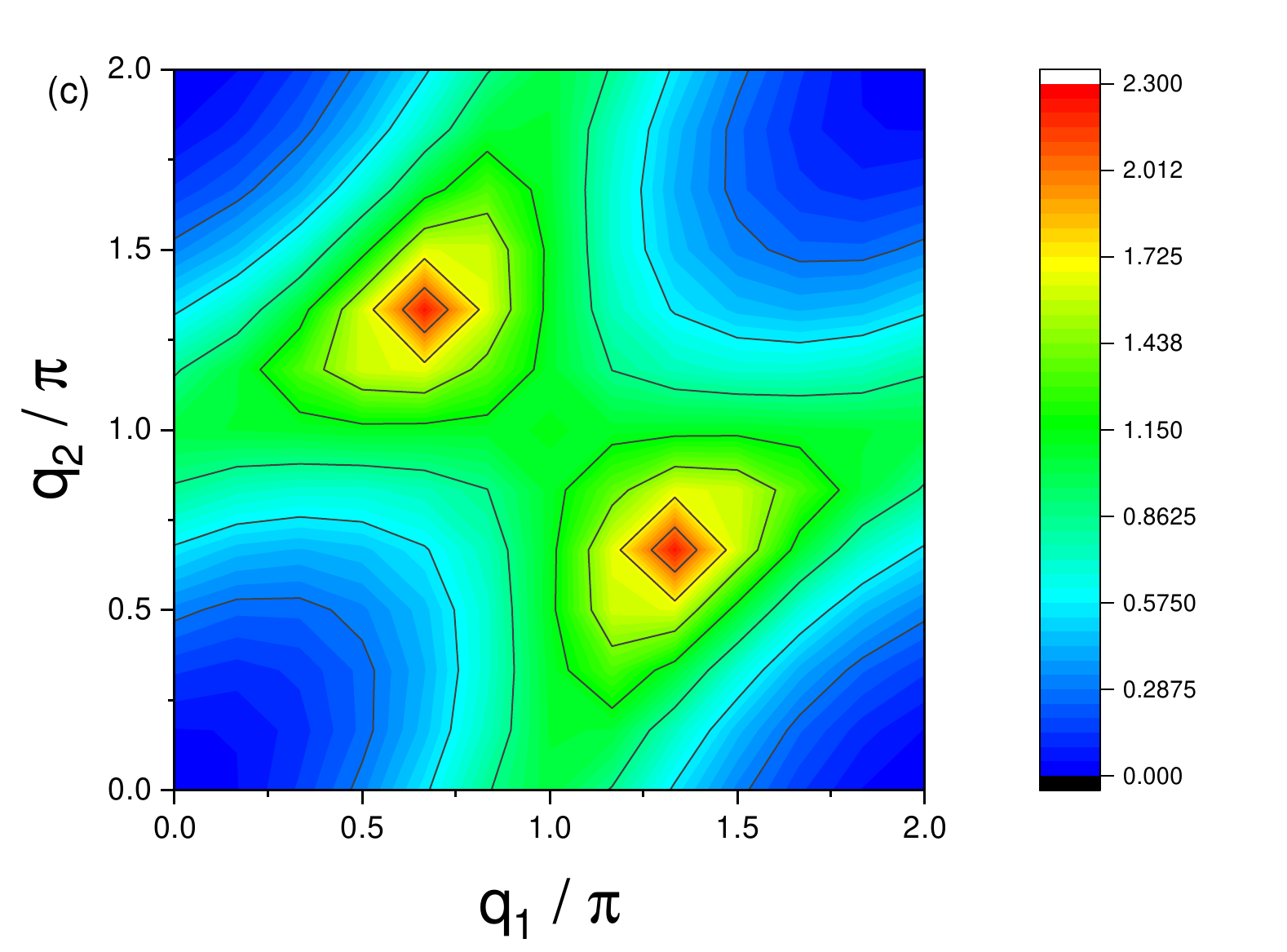}
\includegraphics[width=7cm]{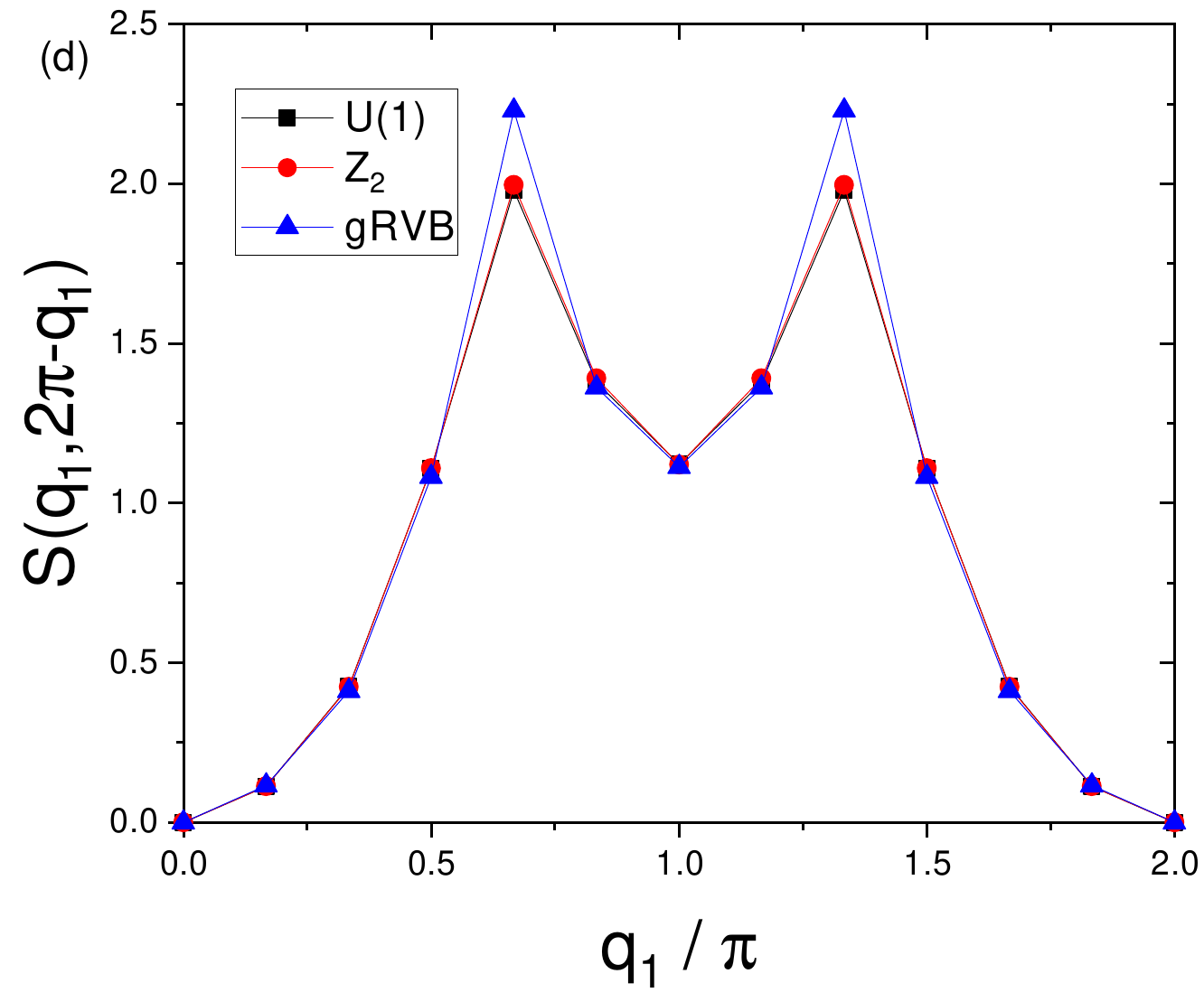}
\caption{The static spin structure factor $S(\mathbf{q})$ of the optimized (a)$U(1)$, (b)$Z_{2}$ and (c)gRVB states for $J_{4}=0$.  $S(\mathbf{q})$ is essentially independent of $J_{4}$ for $0\le J_{4}\le 0.045$. (d)Comparison of the static spin structure factor of the three kind of RVB states along the path $(0,2\pi)\rightarrow(2\pi,0)$.}
\end{figure}

 The energy difference between the general RVB state and the $U(1)$ or the $Z_{2}$ RVB state is more tricky. We find that the optimized gRVB state can not be generated by any short-ranged mean field ansatz. However, the static spin structure factor of the optimized gRVB state is essentially indistinguishable from that of the optimized $U(1)$ or $Z_{2}$ RVB state(see Fig5c and 5d). It is currently impossible to conduct the optimization of the gRVB state on a cluster significantly larger than the $12\times 12$ cluster. For example, on a $24\times 24$ cluster, the number of variational parameters in the gRVB state becomes $\frac{N(N-1)}{2}=165600$, which is too large to be reliably optimized. 

In all, we find that the optimized RVB state in the whole $0\le J_{4}\le 0.045$ regime represents a state essentially equivalent to the $\pi$-flux phase on the triangular lattice, which is the closest counterpart of the 120 degree ordered phase in the space of Fermionic RVB state.  
  
\begin{figure}
\includegraphics[width=8cm]{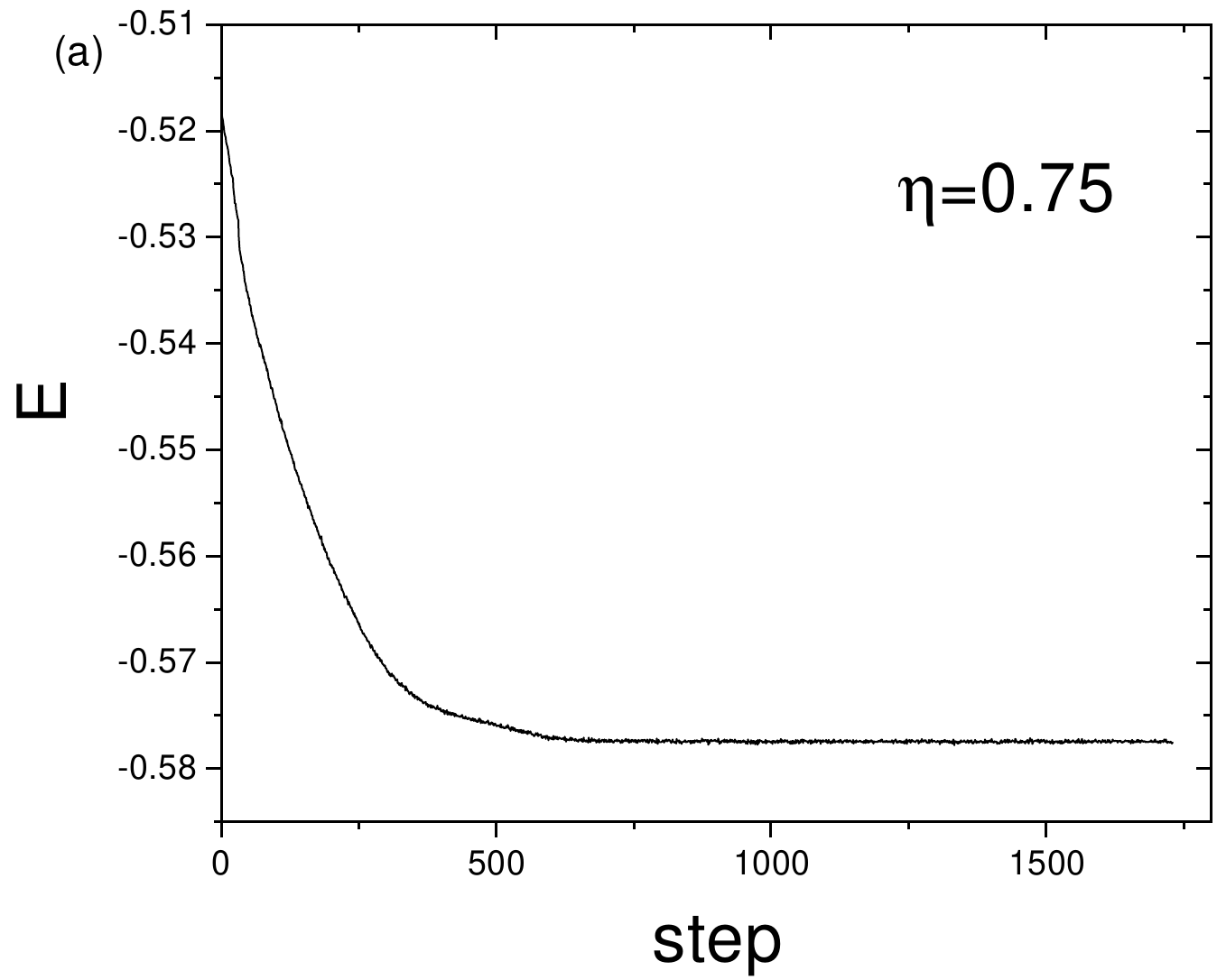}
\includegraphics[width=8cm]{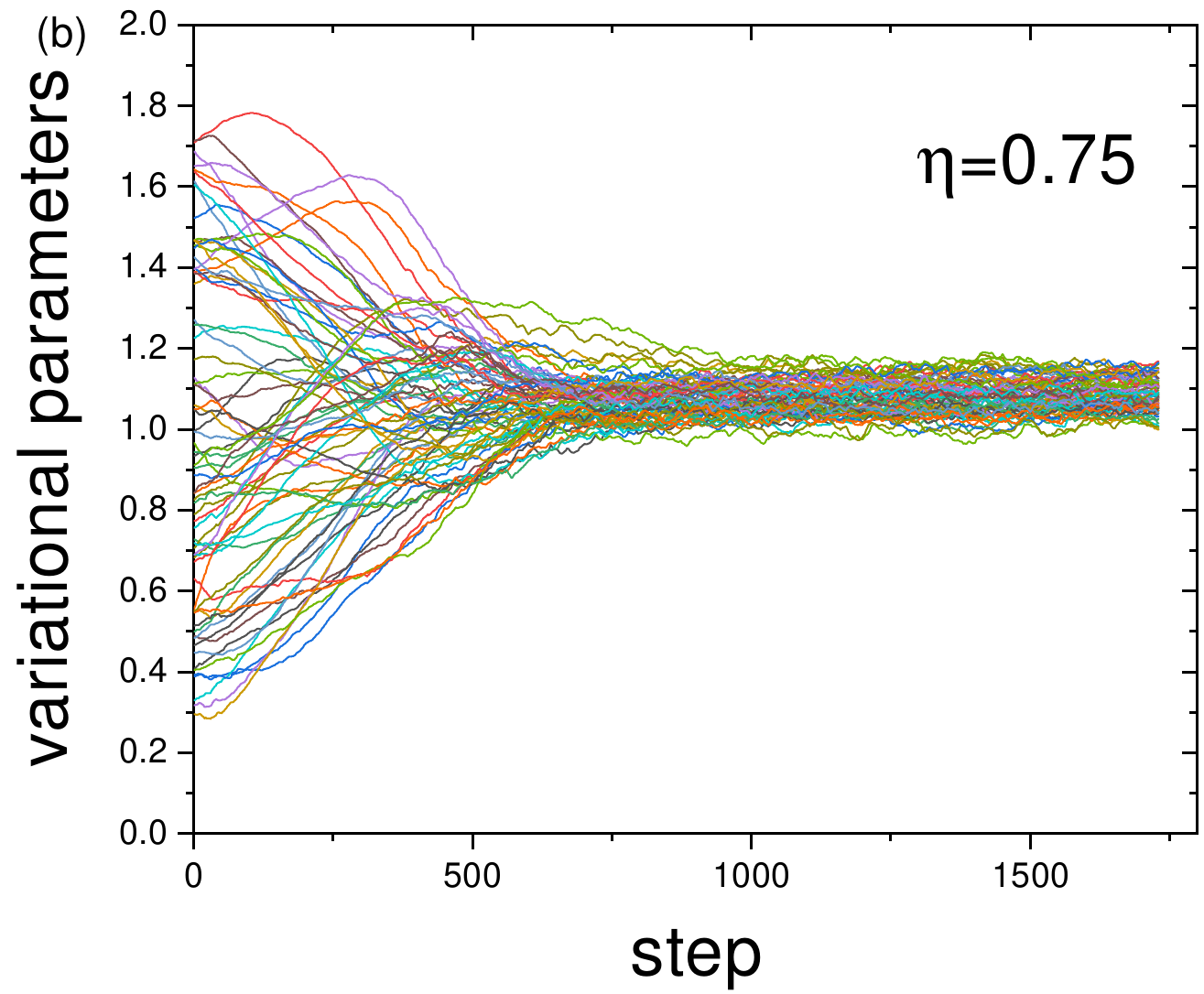}
\caption{The convergence of the variational energy(a) and the variational parameters(b) towards those of the SFS state. Here we set the distance of the initial guess from the SFS ansatz to be $\eta=0.75$.}
\end{figure}

\subsection{The large $J_{4}$ phase}

It is generally believed that in the large $J_{4}$ regime the SFS state is the most favorable variational ground state. Indeed, we find that the SFS state is locally extremely stable in this regime. The SFS state is generated by the following mean field ansatz
\begin{eqnarray}
\chi_{i,j}&=&\Big\{\begin{array}{c}1, \ nearest \ neighbor \\0, \ \ otherwise\end{array} \nonumber\\
\Delta_{i,j}&=&0
\end{eqnarray}
As a result of the translational and rotational symmetry of the mean field ansatz, the energy gradient must be also translational and rotational symmetric. The energy gradient in the SFS phase thus must vanish since the Gutzwiller projected state is invariant under a uniform rescaling of all variational parameters. In other words, the SFS phase is an exact saddle point in the space of Fermionic RVB state. To illustrate the local stability of the SFS state, we have performed variational optimization at $J_{4}=0.15$ starting from the following initial guess of the mean field ansatz
\begin{eqnarray}
\chi_{i,j}&=&\Big\{\begin{array}{c}1\pm \eta r, \ nearest \ neighbor \\0, \ \ otherwise\end{array} \nonumber\\
\Delta_{i,j}&=&0
\end{eqnarray}
Here $r$ is a random number distributed uniformly in the range of $r\in(0,1)$, $\eta$ is a constant measuring the distance of the initial guess from the SFS ansatz. As is shown in Fig.6, even if we choose $\eta$ as large as $\eta=0.75$, the variational energy still converges to a value very close to that of the SFS state, which is $E\approx-0.576$ at $J_{4}=0.15$. The corresponding variational parameters also converge to that of the SFS ansatz.

However, the SFS state is only locally stable. To illustrate this point, we have performed variational optimization starting from fully random initial guess of the variational parameters. The optimization is done for $J_{4}=0.15$ on a $12\times12$ cluster with periodic boundary condition. A preliminary trial shows that the optimized variational ground state exhibits an approximate $4\times6$ modulation in its local spin correlation pattern. This is illustrated in Fig.7. We note that the modulation in the local spin correlation is very strong, ranging from almost pure spin singlet correlation to pure spin triplet correlation between nearest neighboring spins. The translational symmetry is seriously broken. 

We then refine the optimization by assuming a $4\times6$ periodicity in the variational parameters. We find that for both the $U(1)$ and the $Z_{2}$ mean field ansatz, the optimized variational energy converge to values much lower than that of the SFS state. Fig.8 and Fig.9 illustrate the convergence of the variational energies for randomly chosen initial guess of the $U(1)$ and the $Z_{2}$ mean field ansatz. For $U(1)$ mean field ansatz, we find that about $1/3$ of the variational optimization trials converge to the lowest variational energy of $E\approx-0.603$. For $Z_{2}$ mean field ansatz, about $1/2$ of the variational optimization trials converge to the lowest variational energy of $E\approx-0.606$. The optimized variational energy for both types of RVB states are about $5\%$ lower than the energy of the SFS state, which is $E\approx-0.576$ at $J_{4}=0.15$. 

\begin{figure}
\includegraphics[width=9cm]{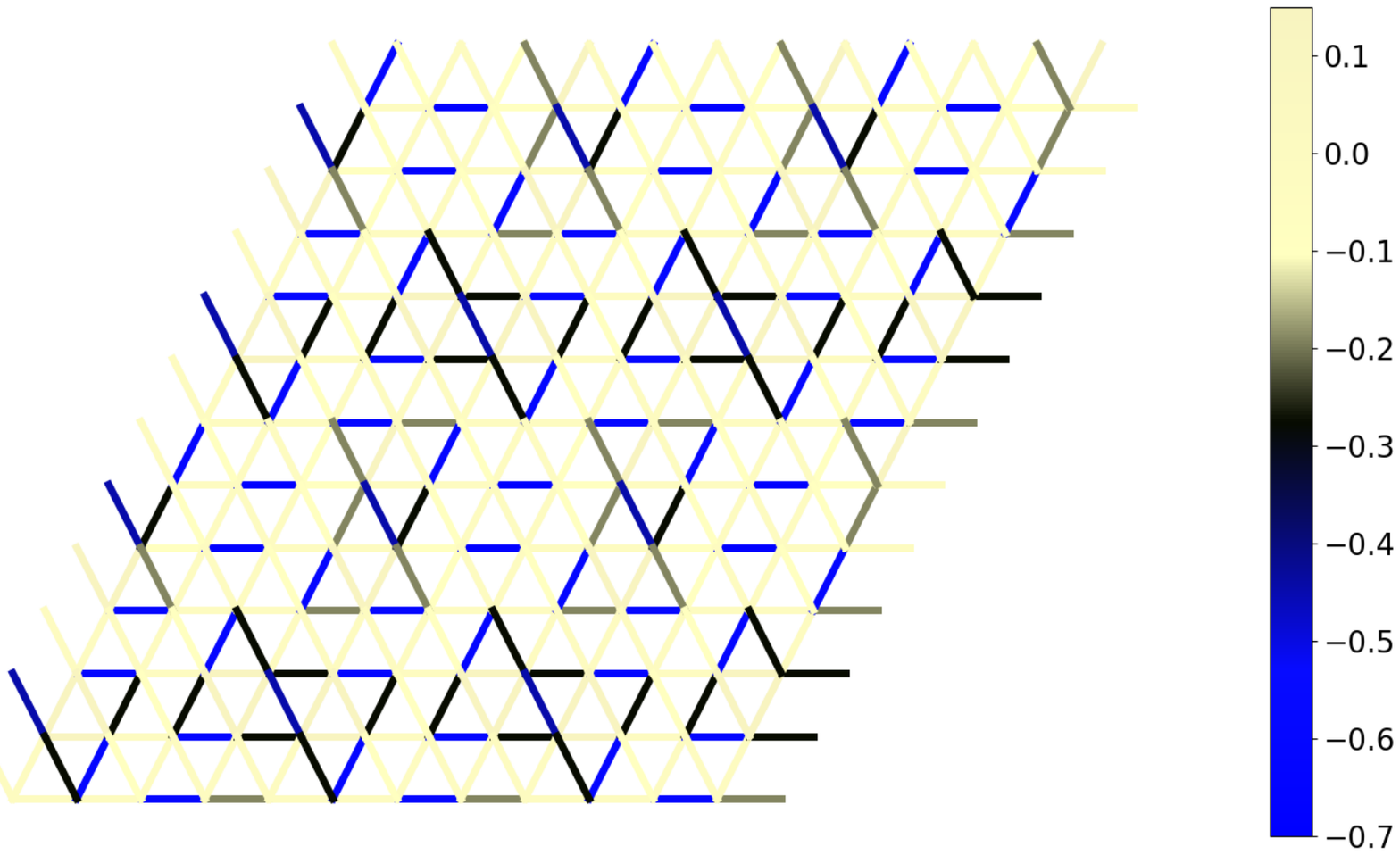}
\caption{The optimized variational ground state at $J_{4}=0.15$ exhibits $4\times6$ modulation in its local spin correlation pattern. Shown here is the spin correlation between nearest neighboring sites. Note that the magnitude of the modulation in the local spin correlation is very large, ranging from nearly pure spin singlet correlation to almost pure spin triplet correlation between nearest neighboring spins.}
\end{figure}  

\begin{figure}
\includegraphics[width=8cm]{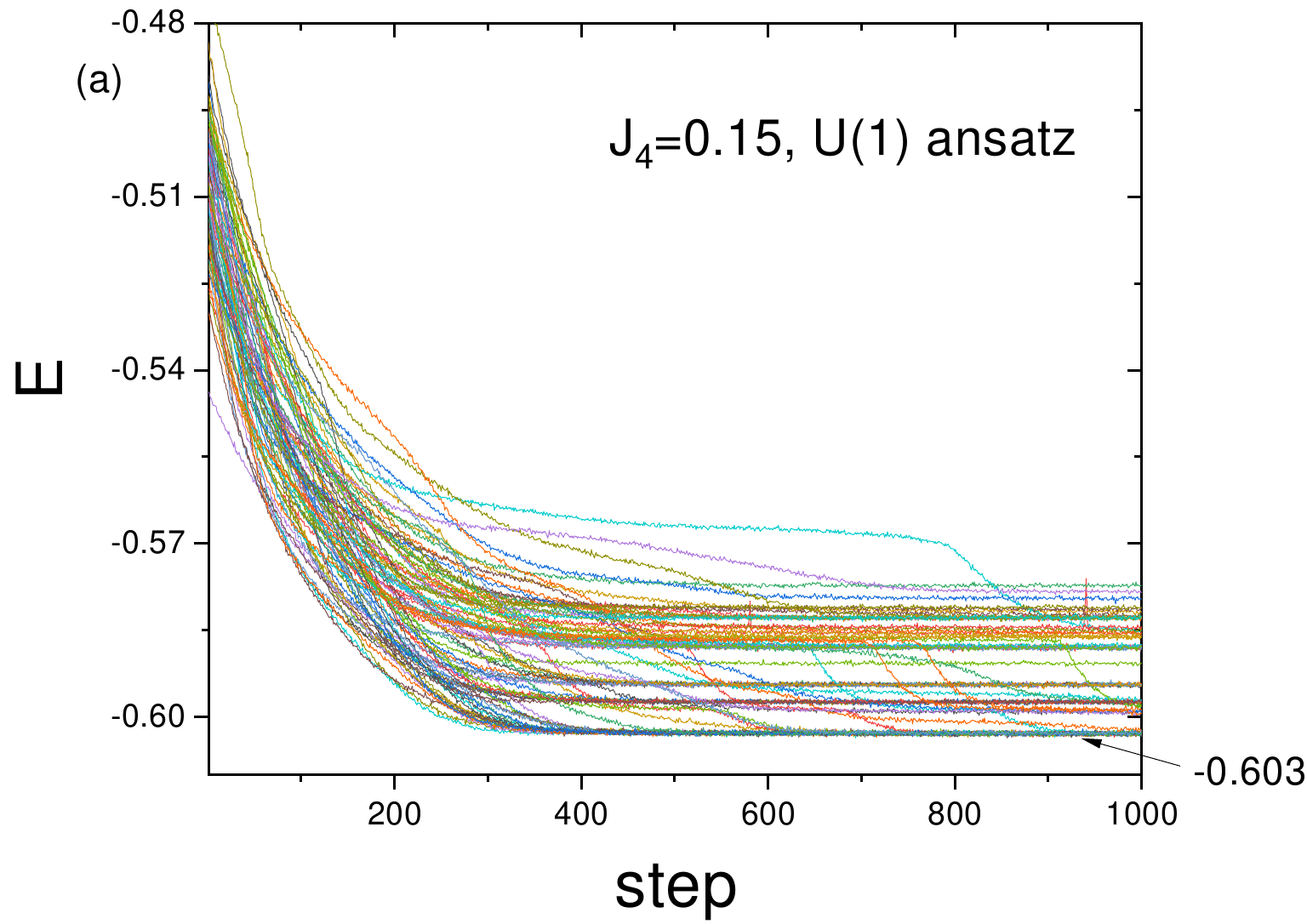}
\includegraphics[width=6.5cm]{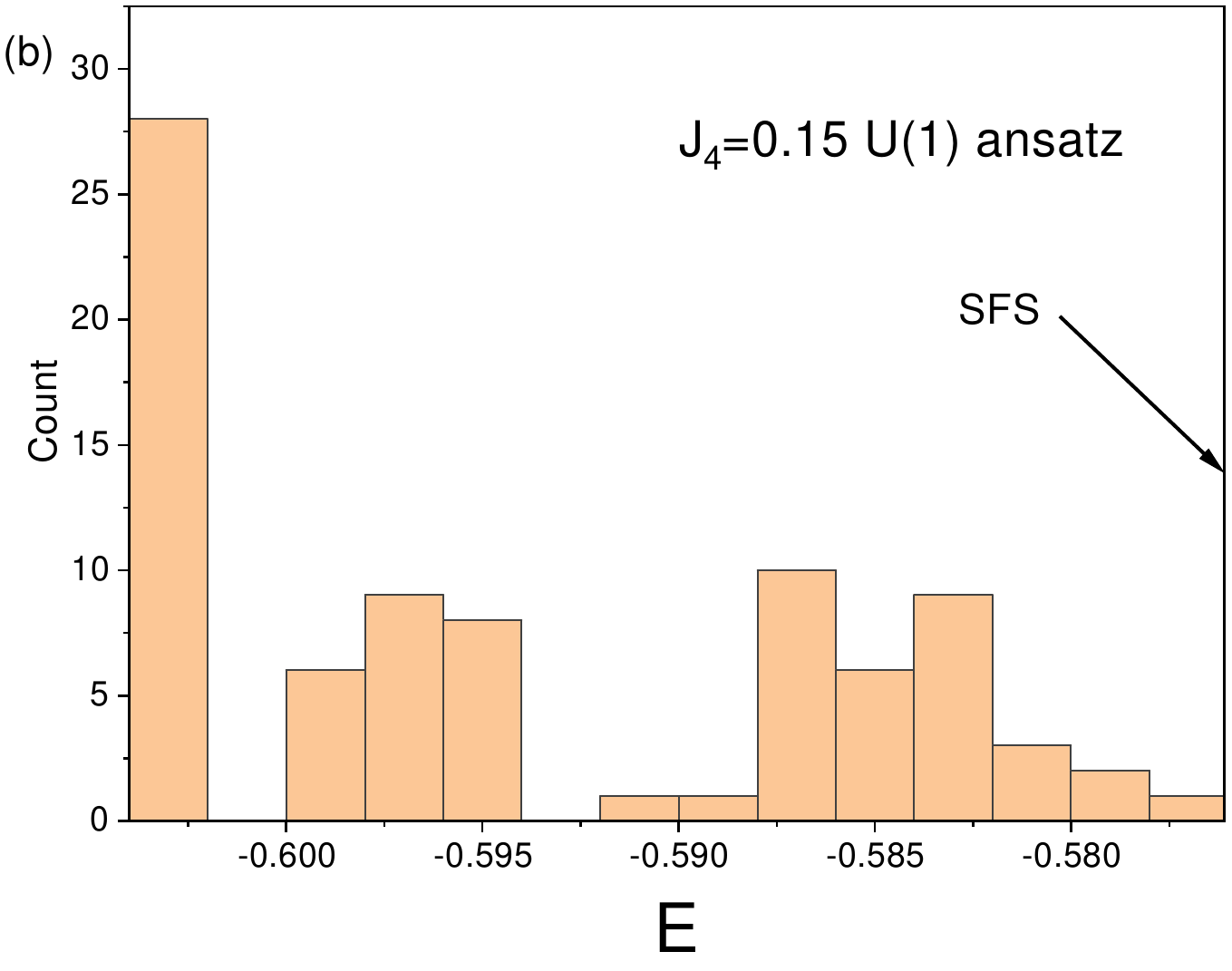}
\caption{(a)The convergence of the variational energy at $J_{4}=0.15$ for a $U(1)$ mean field ansatz. Shown here is the results from 82 optimization trials starting from randomly chosen initial guess of the $U(1)$ mean field ansatz with $4\times6$ periodicity. The calculation is done on a $12\times12$ cluster with periodic boundary condition. About $1/3$ of the optimization trials converge to the lowest variational energy of $E\approx-0.603$. This is much lower than the energy of the SFS state, which is $E\approx-0.576$. (b)The histogram of the optimized variational energies from the 82 optimization trials.}
\end{figure}  

\begin{figure}
\includegraphics[width=8cm]{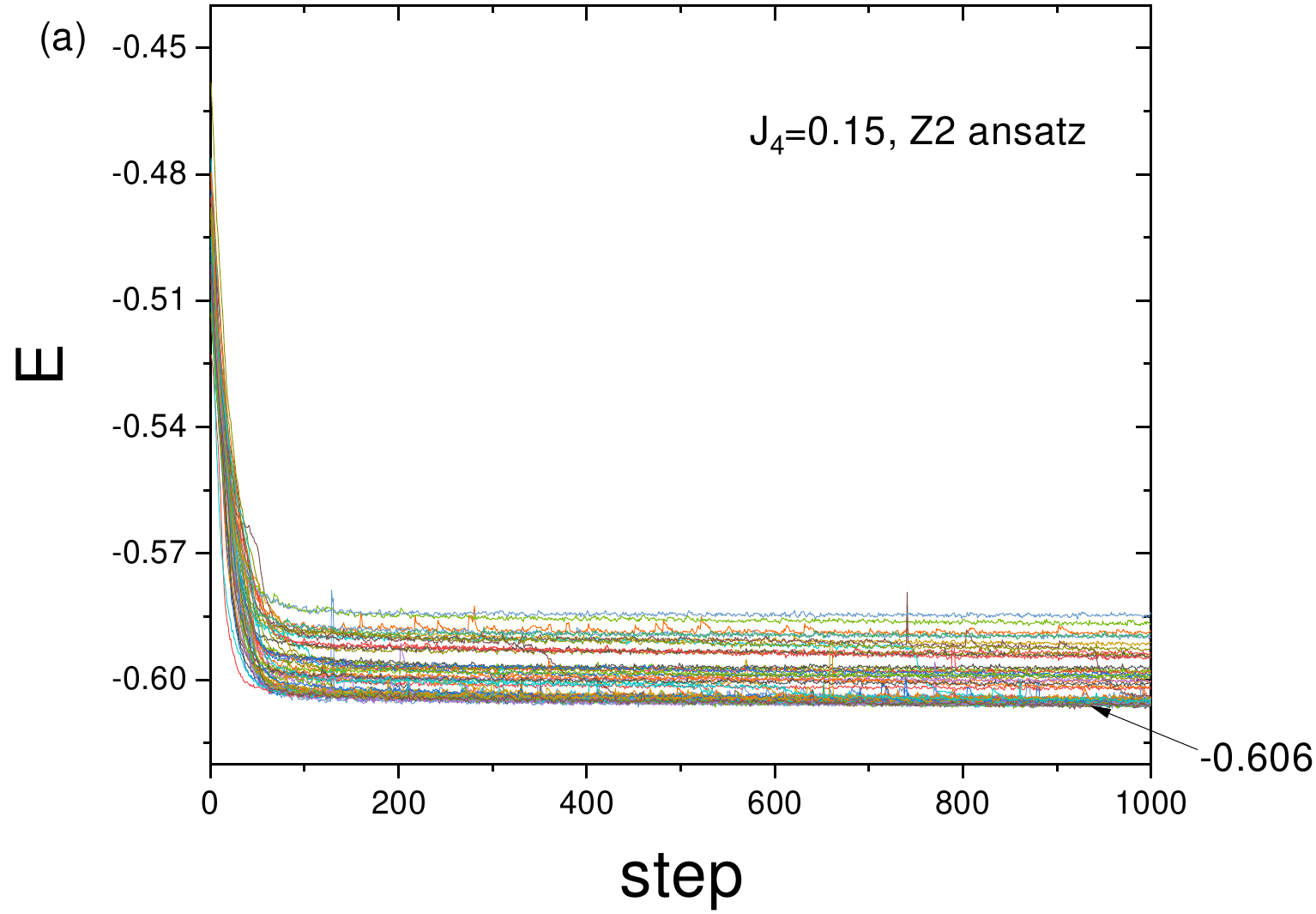}
\includegraphics[width=6.5cm]{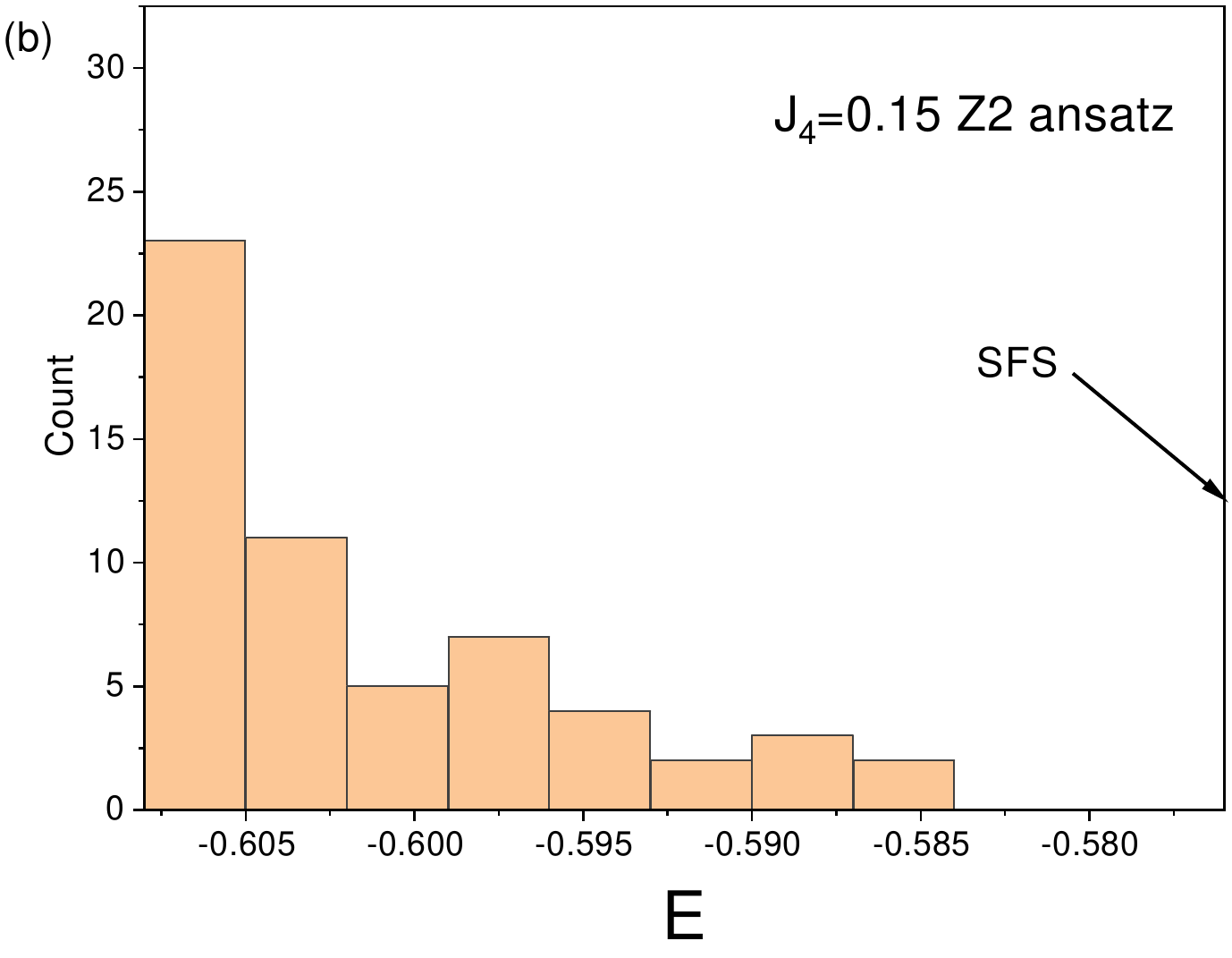}
\caption{(a)The convergence of the variational energy at $J_{4}=0.15$ for a $Z_{2}$ mean field ansatz. Shown here is the results from 57 optimization trials starting from randomly chosen initial guess of the $Z_{2}$ mean field ansatz with $4\times6$ periodicity. The calculation is done on a $12\times12$ cluster with periodic boundary condition. About $1/2$ of the optimization trials converge to the lowest variational energy of $E\approx-0.606$. This is also much lower than the energy of the SFS state, which is $E\approx-0.576$. (b)The histogram of the optimized variational energies from the 57 optimization trials.}
\end{figure}  

On the $12\times12$ cluster studied here, the most general periodicity of the mean field ansatz that is compatible with the periodic boundary condition of the cluster is $C_{1}\times C_{2}$, in which $C_{1,2}=1,2,3,4,6,12$ denotes the period in the $\mathbf{a}_{1}$ and $\mathbf{a}_{2}$ direction. We have performed refined variational optimization assuming each of such periodicity. We find that the lowest variational energy is always obtained with the $4\times6$ periodicity at $J_{4}=0.15$.

We find that the optimized variational energy as a function of $J_{4}$ exhibits a very small curvature in the large $J_{4}$ regime(see Fig.10). This implies that the variational ground state is almost independent of $J_{4}$ in this part of the phase diagram also.  A naive linear extrapolation of the variational energy implies that the SFS state can not be the true ground state of the $J_{1}-J_{4}$ model for $J_{4}\le 0.5$. To check if the SFS state can be stabilized at still larger $J_{4}$, we have performed variational optimization for $J_{4}=10$, a value that is too large to be realistic for real materials. We find that a symmetry breaking phase with the same $4\times 6$ periodicity in its spin correlation pattern is still significantly lower in energy than the SFS state. More specifically, we find that the optimized variational energy is $E\approx-16.09$ at $J_{4}=10$. This is again about $5\%$ lower than the energy of the SFS state, which is $E\approx-15.33$ at $J_{4}=10$. Fig.11 illustrates the local spin correlation pattern in the optimized state, which is similar to that at $J_{4}=0.15$. We thus conclude that the SFS state is never the best RVB state for the $J_{1}-J_{4}$ model and that the $4\times 6$ state is the best variational ground state in the whole range of $J_{4}\ge 0.09$. Such a state breaks both the translational and the point group symmetry of the model and exhibits $4\times6$ modulation in its local spin correlation pattern.

\begin{figure}
\includegraphics[width=8cm]{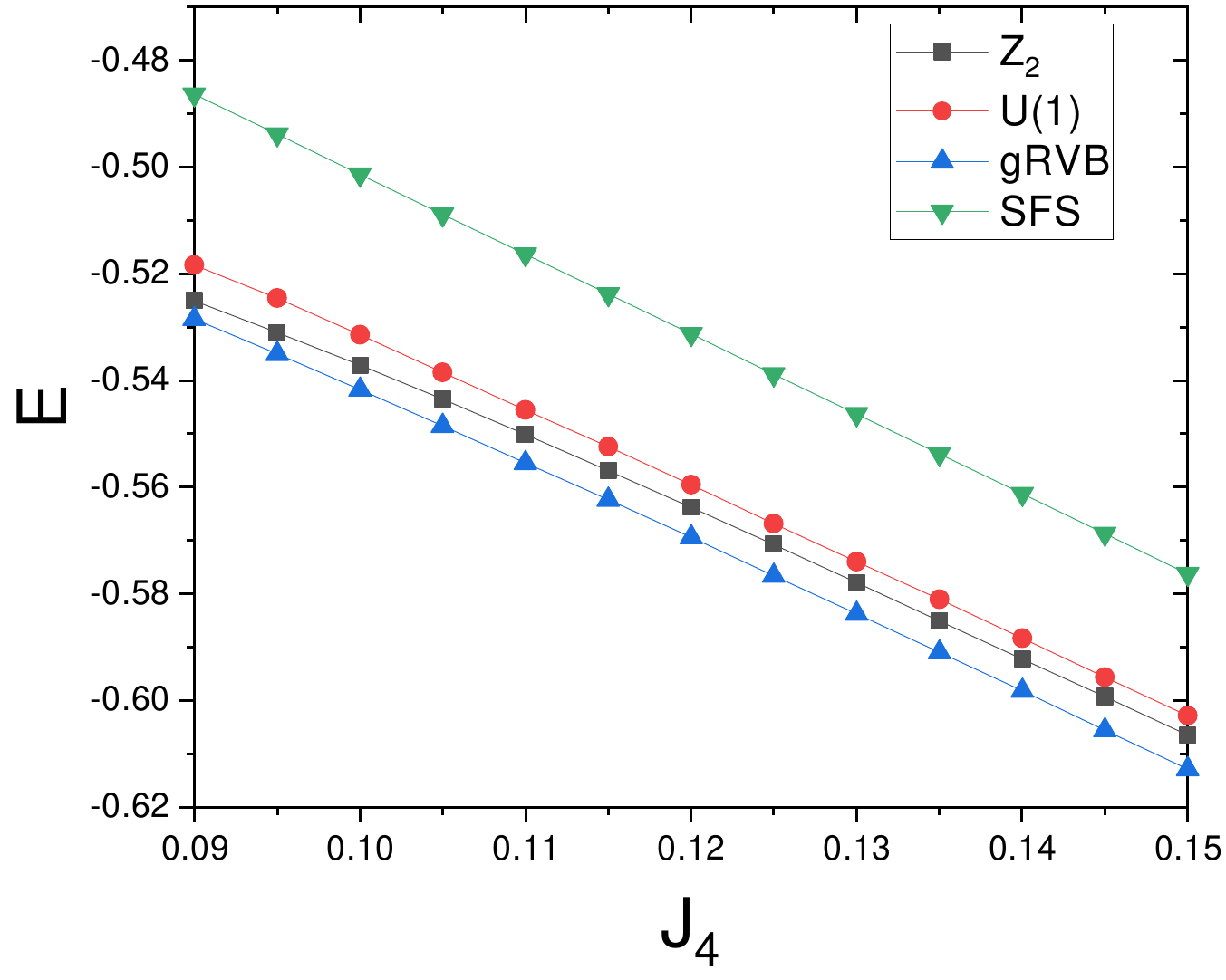}
\caption{The variational energies of the three kinds of RVB state in the range of $0.09\le J_{4}\le 0.15$. The computation is done on a $12\times 12$ cluster with periodic boundary condition. The optimized variational ground state is found to be almost independent of $J_{4}$ in this regime, as is evident from the smallness of the curvature in the optimized variational energy. Such a state is found to exhibit a $4\times6$ modulation in its local spin correlation pattern.}
\end{figure}

\begin{figure}
\includegraphics[width=9cm]{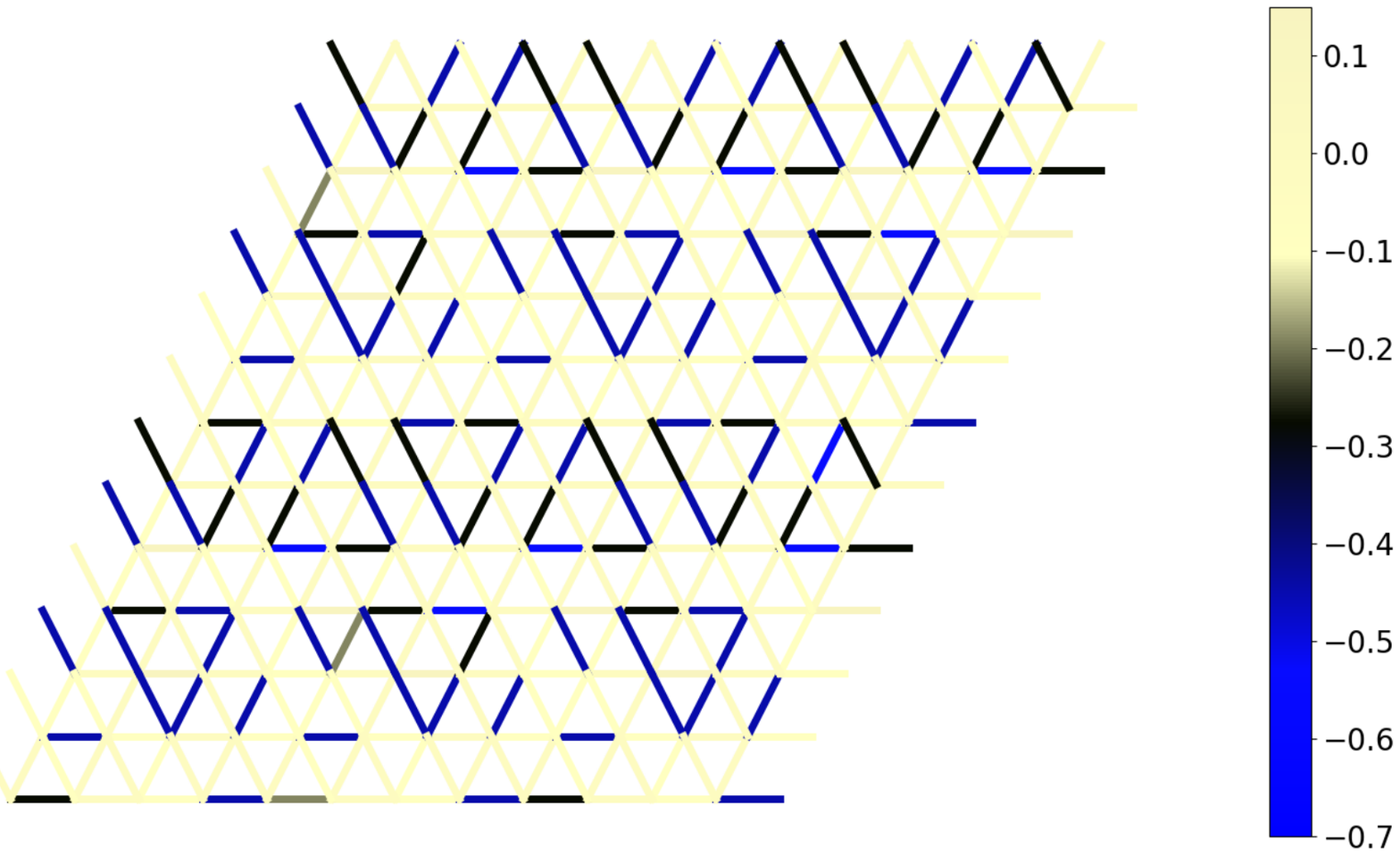}
\caption{The optimized variational ground state at $J_{4}=10$ exhibits the same $4\times6$ modulation in its local spin correlation pattern as that for $J_{4}=0.15$. Shown here is the spin correlation between nearest neighboring sites.}
\end{figure}  

\begin{figure}
\includegraphics[width=8.5cm]{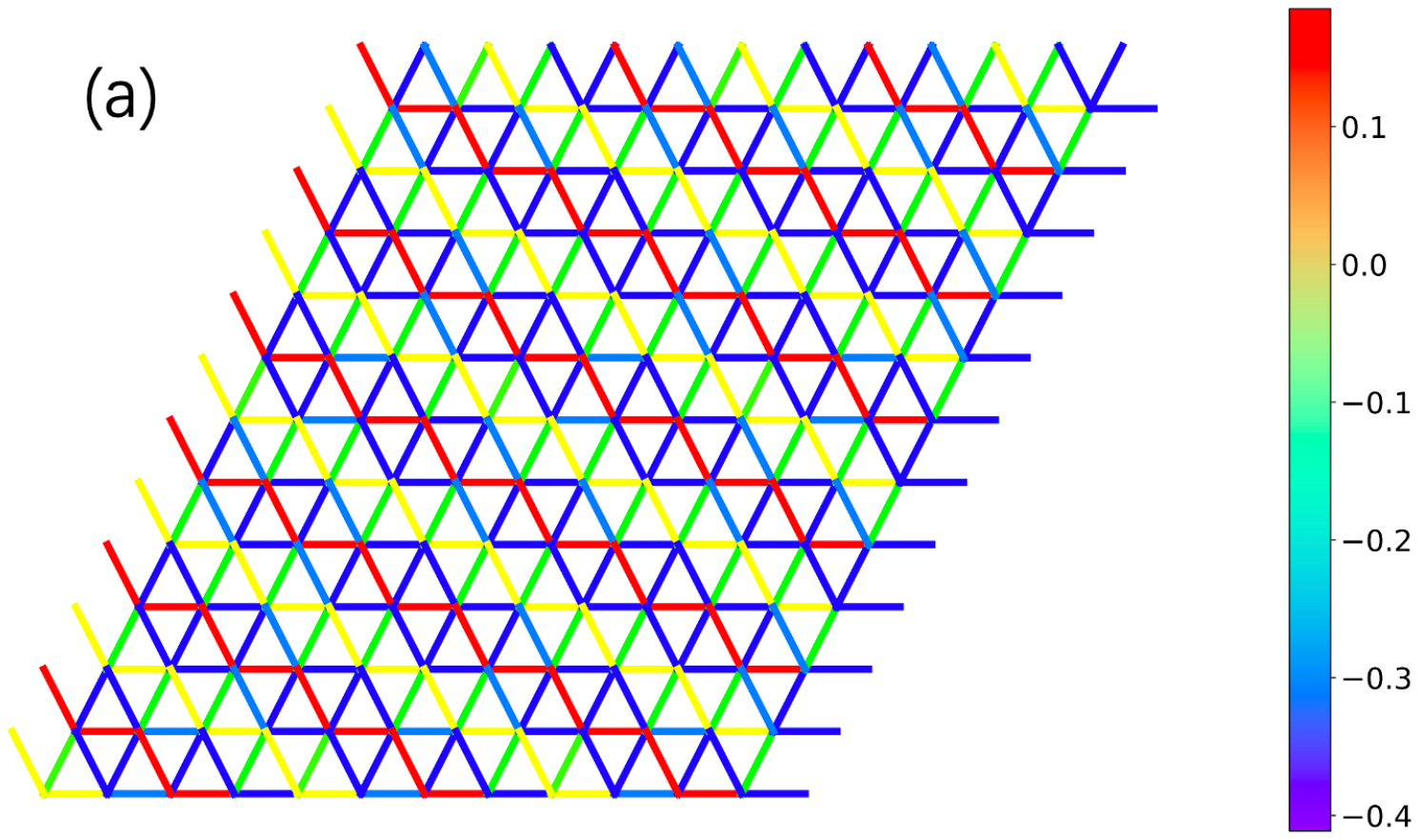}
\includegraphics[width=8.5cm]{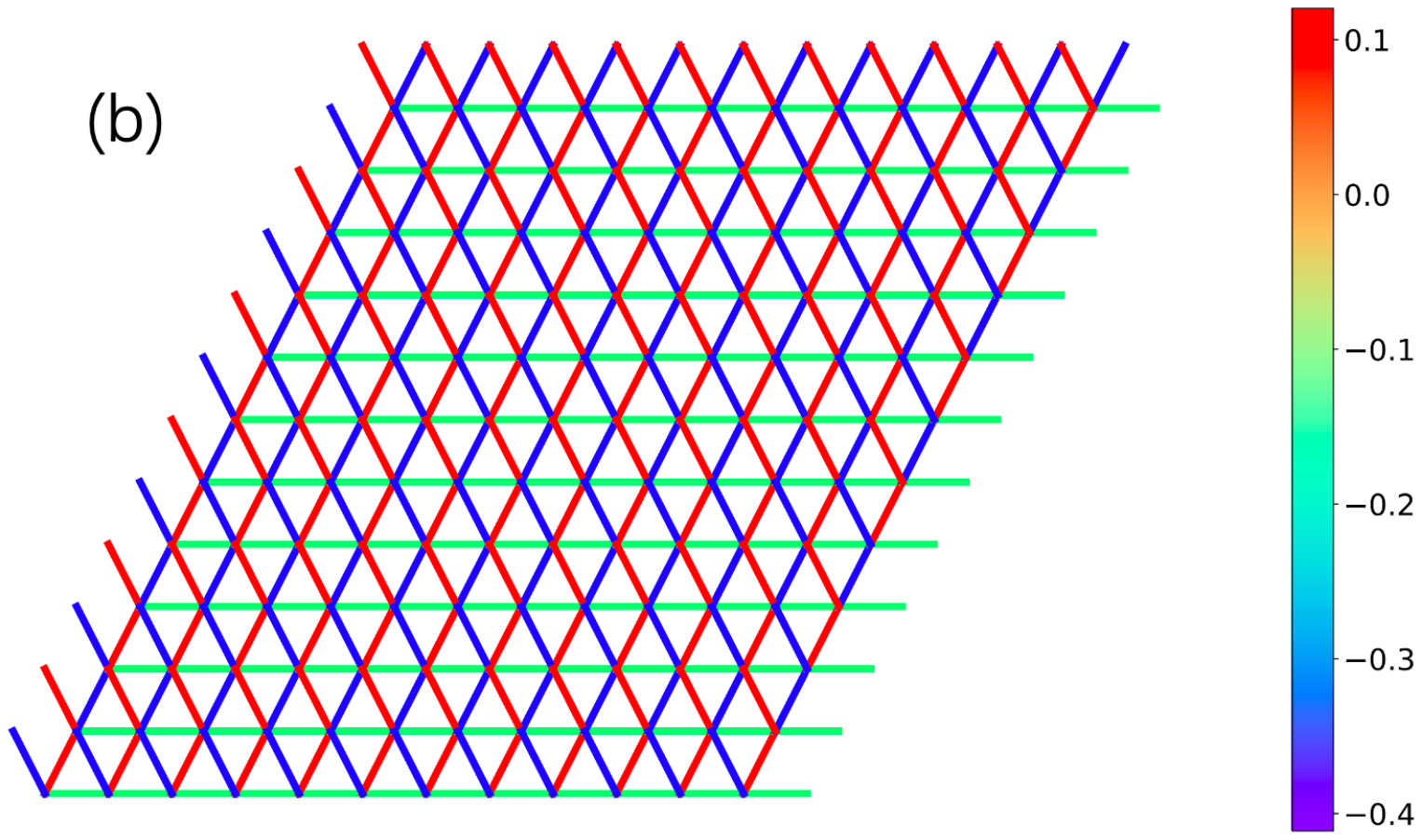}
\includegraphics[width=8.5cm]{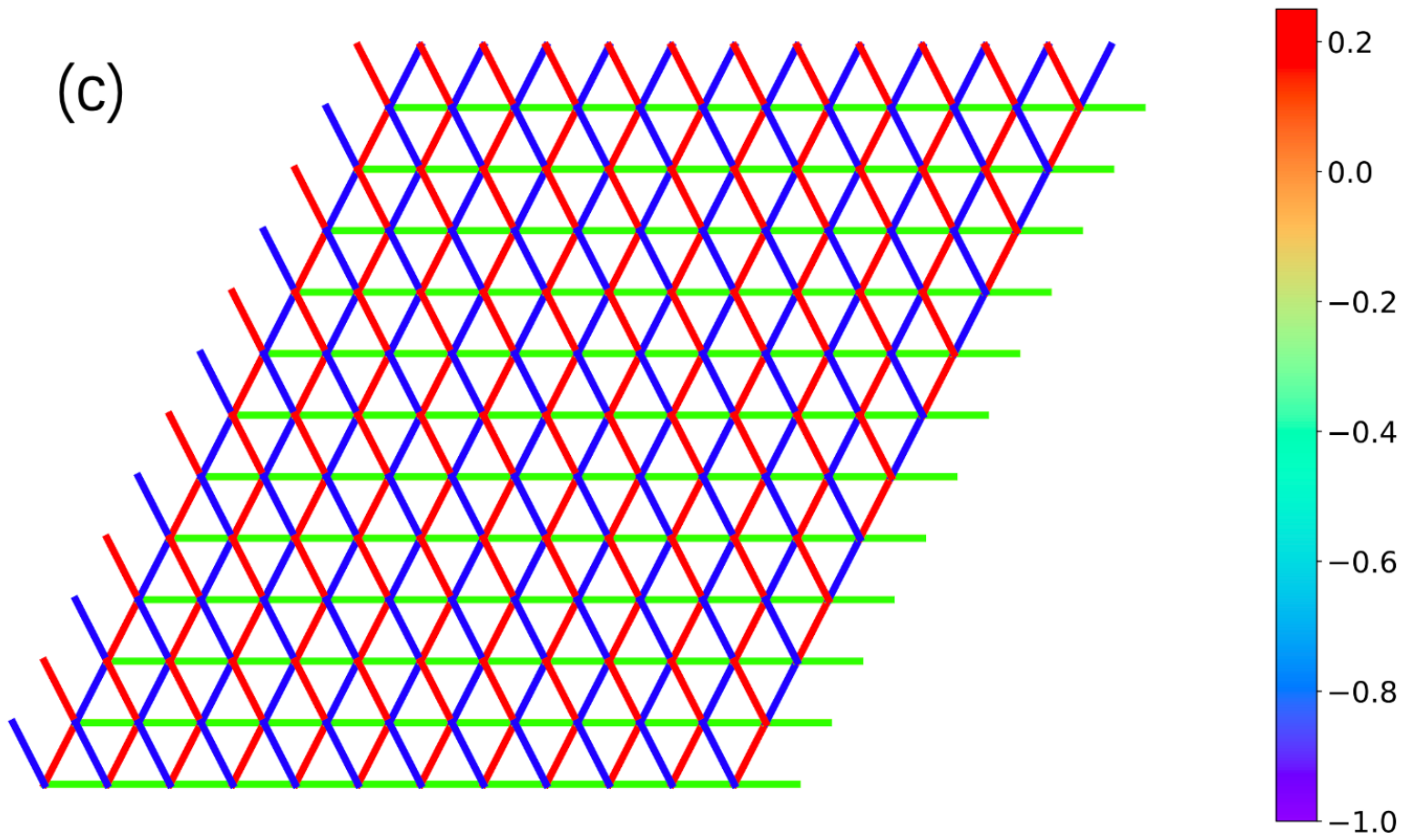}
\caption{The local spin correlation pattern of the optimized variational ground state for (a)$0.045\le J_{4}\le 0.055$ and (b)$0.055\le J_{4}\le 0.09$. Both states exhibit zigzag spin correlation pattern, with the antiferromagnetic correlated backbones(plotted here in blue) running through (a)the $2\mathbf{a}_{1}-\mathbf{a}_{2}$ direction and (b)the $2\mathbf{a}_{2}-\mathbf{a}_{1}$ direction. (c)The local spin correlation pattern in a classical zigzag state, in which the neighboring spin correlation takes the value of $-1$, $+\frac{1}{4}$ and $-\frac{1}{4}$ on the blue, red and green bonds. The zigzag pattern shown in (a) differs from that shown in (b) by an additional two-fold translational symmetry breaking.}
\end{figure}  

\subsection{The intermediate phase(s)}
The optimized variational ground state in the intermediate regime of $0.045\le J_{4}\le 0.09$ is characterized by a zigzag modulation in its local spin correlation pattern. As is illustrated in Fig.12, the zigzag pattern manifests itself most evidently in the antiferromagnetic correlated backbones running through the $2\mathbf{a}_{1}-\mathbf{a}_{2}$ or equivalent directions of the triangular lattice. The translational symmetry and the rotational symmetry are thus spontaneously broken. Looking more closely, we find that the zigzag pattern in the $0.045\le J_{4}\le 0.055$ regime differs from that in the $0.055\le J_{4}\le 0.09$ regime by an additional two-fold translational symmetry breaking.
Fig.13 shows the optimized variational energy for the zigzag phase. The energy difference between the SFS state and the zigzag state is even more significant than that in the $4\times6$ phase. A closer inspection also shows that the energy difference between the $U(1)$ and the $Z_{2}$ state is more significant in the zigzag regime than that in the $\pi$-flux and the large $J_{4}$ regime.
\begin{figure}
\includegraphics[width=8cm]{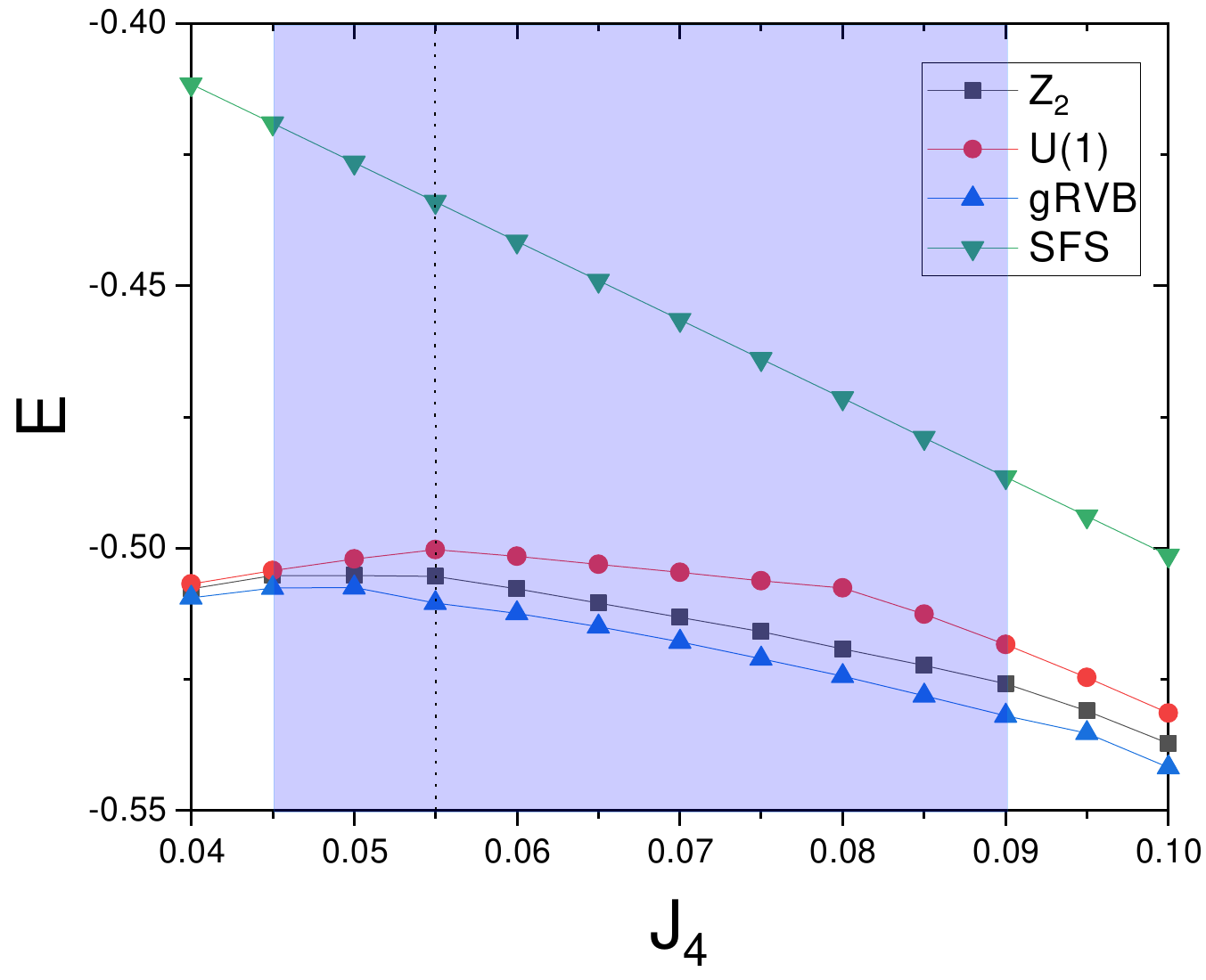}
\caption{The variational energies of the three kinds of RVB state as functions of $J_{4}$ in the range of $0.045\le J_{4}\le 0.09$. The computation is done on a $12\times12$ cluster with periodic boundary condition. The optimized variational ground state in this regime is found to exhibit a zigzag local spin correlation pattern. The zigzag pattern in the $0.045\le J_{4}\le 0.055$ regime differs from that in the $0.055\le J_{4}\le 0.09$ regime by an additional two-fold translational symmetry breaking.}
\end{figure}  

Instead of the zigzag phase found here, several translational invariant phases have been proposed in the intermediate $J_{4}$ regime in the literature. In Ref.[\onlinecite{Grover}], a nodal $d$-wave $Z_{2}$ state with nematic spinon dispersion is proposed to be the variational ground state in the intermediate regime. It is later found that another $Z_{2}$ state with $d+id$-wave spinon pairing and quadratic band touching(QBT) in the spinon dispersion is more favorable than the nodal d-wave state in a tiny range of $J_{4}$\cite{Xu}. The QBT state hosts gapless spinon excitation with a finite density of state and a gapped gauge fluctuation spectrum as a result of the $d+id$-wave spinon pairing. Such an excitation characteristics is argued to be helpful to resolve the puzzle related to the anomalous gauge fluctuation correction to the specific heat in the SFS state. While we think that the anomalous gauge fluctuation correction in the SFS state is only a theoretical artifact of the conventional gauge theory argument\cite{Li2}, it is nevertheless interesting to compare the variational energies of these novel states to our variational result. In Fig.14, we plot the variational energies of the major candidates of the variational ground state of the $J_{1}-J_{4}$ model in the intermediate $J_{4}$ regime. From this figure we see that the energy advantage of the QBT state over the nodal $d$-wave state is meaningless when compared to the huge energy difference between these states and the zigzag state we find from variational optimization. This establishes firmly the zigzag nature of the variational ground state in the intermediate regime.   

A zigzag phase has also been reported in a similar parameter regime in a recent DMRG study of the $J_{1}-J_{2}-J_{4}$ model\cite{Moore}, in which $J_{2}$ denotes the exchange coupling between next nearest neighboring spins. Different from what we found here, the zigzag phase reported in Ref.[\onlinecite{Moore}] exhibits magnetic long range order. The Fermionic RVB state we adopted in this study is not expected to describe accurately such magnetic long range ordered phases. However, as we have seen in the case of the $\pi$-flux phase for $0\leq J_{4}\leq 0.045$, the Fermionic RVB state can nevertheless reproduce correctly the qualitative feature of the spin structure factor of the magnetic ordered phase. We find that this is also the case in the zigzag phase. In Fig.15 we plot the spin structure factor of the model at $J_{4}=0.06$ calculated from the optimized gRVB state\cite{note1}. The spin structure factor is characterized by the prominent peaks at $\mathbf{q}=(\pm\frac{\pi}{2},\pi)$ and the weaker peak at $\mathbf{q}=(\pi,0)$. These are exactly the positions of the magnetic Bragg peaks in the classical zigzag phase(illustrated in Fig15c). 

With these considerations in mind, it is better to interpret the $\pi$-flux phase in the small $J_{4}$ regime and the zigzag phase in the intermediate $J_{4}$ regime both as the closest approximation of the corresponding magnetic ordered phases, namely the 120 degree ordered phase and the zigzag ordered phase. Thus, the transition at $J_{4}=0.045$ should be better understood as a first order transition between two magnetic ordered phases, rather than the transition between a spin liquid phase and a valence bond solid phase. According to Ref.[\onlinecite{Xu}], the 120 degree ordered phase becomes degenerate with the nodal d-wave state at $J_{4}=0.0525$. Since the energy of the zigzag phase found here is lower than that of the nodal d-wave state at $J_{4}=0.0525$, we expect that the transition between the 120 degree ordered phase and the zigzag phase to occur at a smaller value of $J_{4}$ than $0.0525$. This is indeed the case in our calculation.

\begin{figure}
\includegraphics[width=8.5cm]{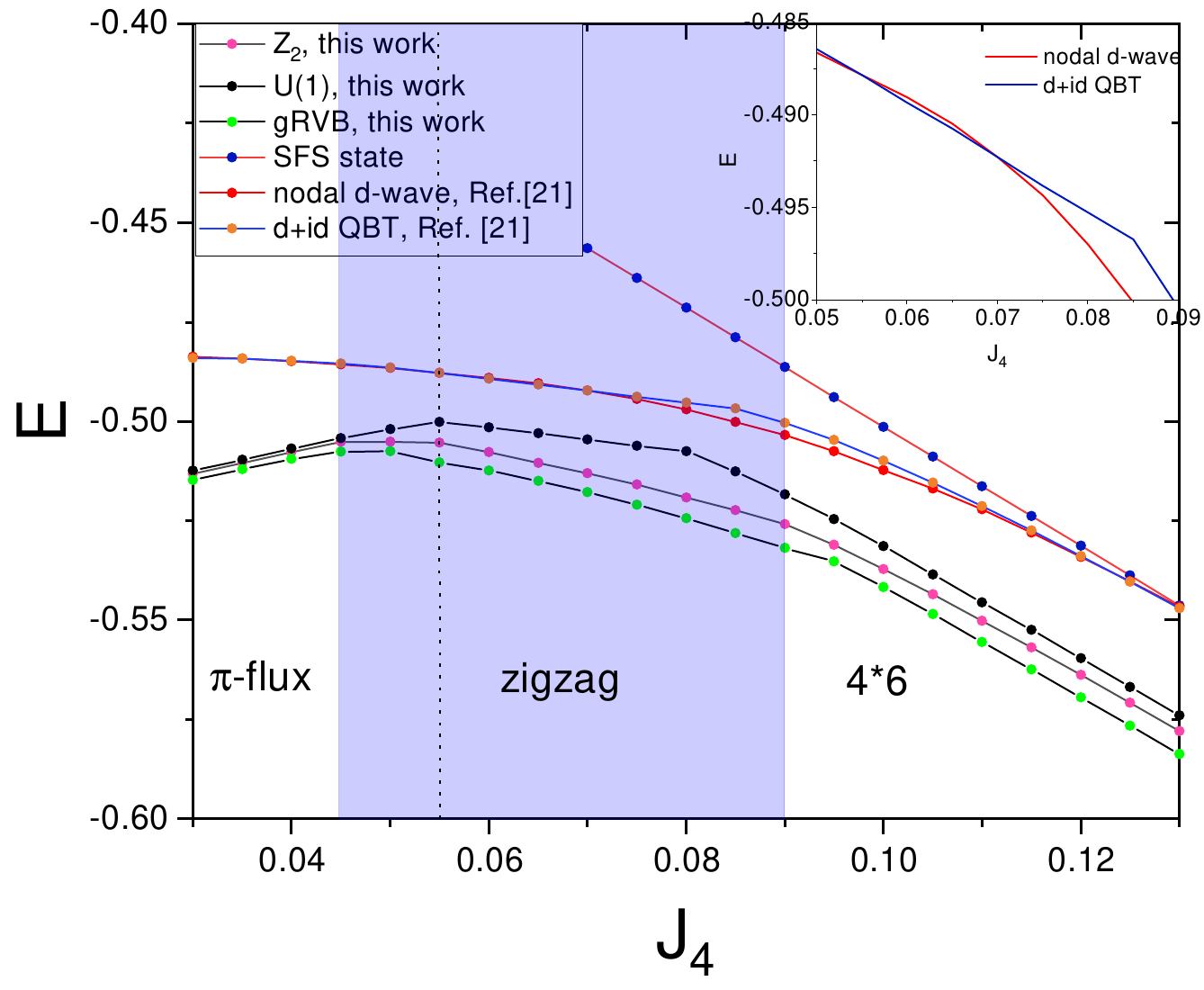}
\caption{The variational energies of the major candidates of the variational ground state of the $J_{1}-J_{4}$ model in the intermediate regime. The inset shows an enlarged view of the crossing region of the nodal $d$-wave and the $d+id$ QBT state. The energy of the nodal $d$-wave and the QBT state are taken from Ref.[\onlinecite{Xu}]. The results of this work are computed on a $12\times12$ cluster with periodic boundary condition.}
\end{figure}

\begin{figure}
\includegraphics[width=8.5cm]{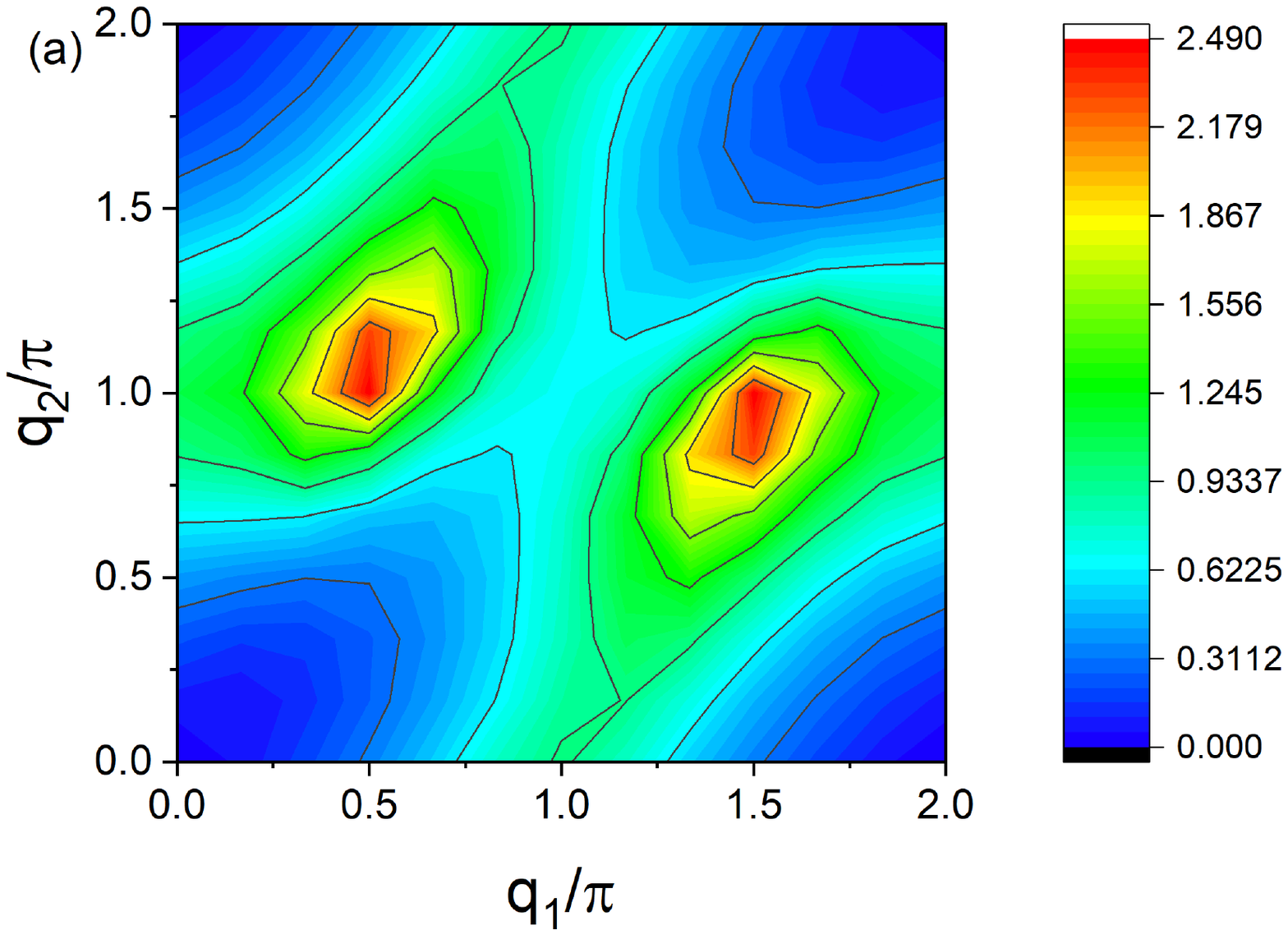}
\includegraphics[width=8.5cm]{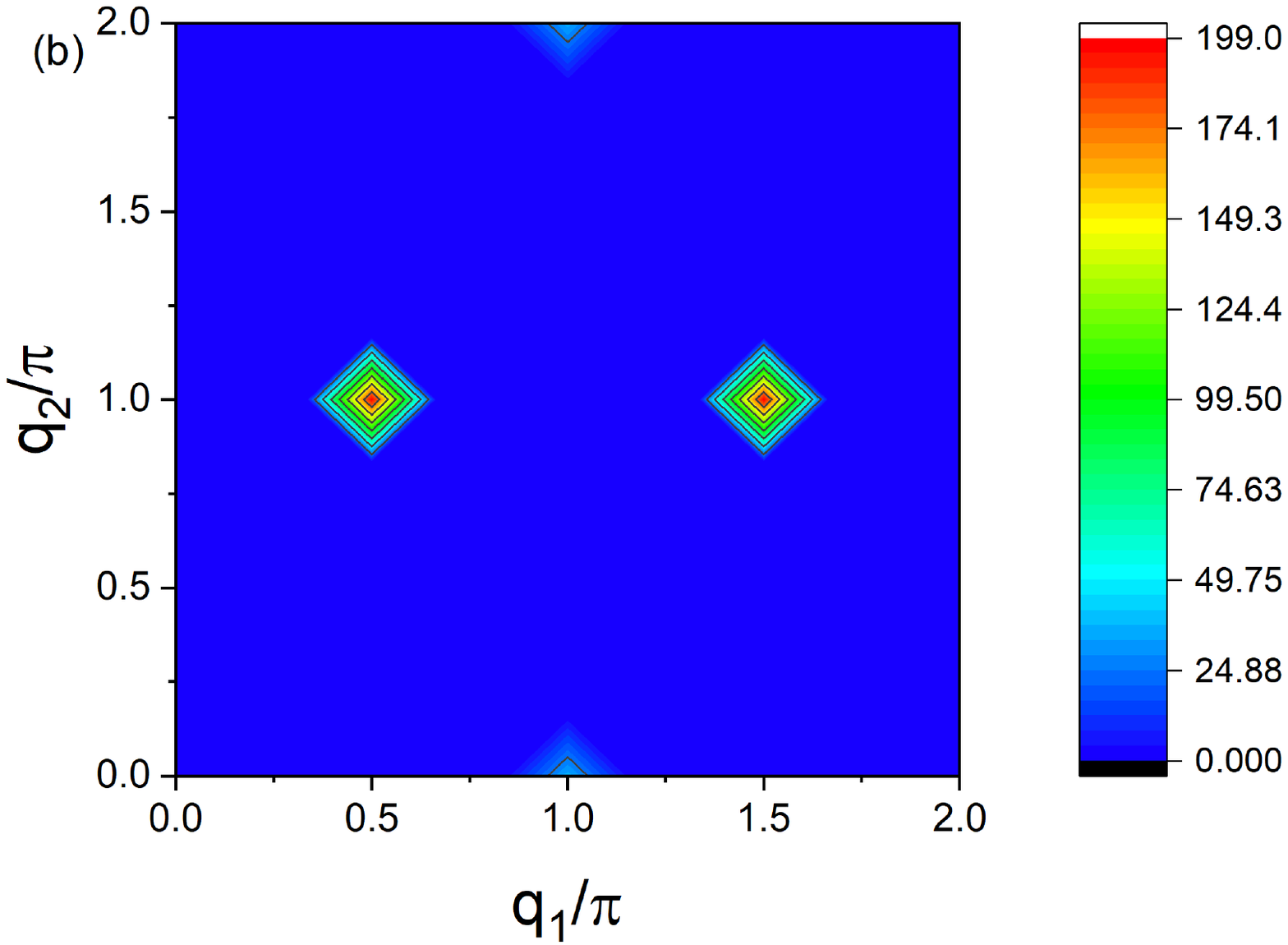}
\includegraphics[width=8.5cm]{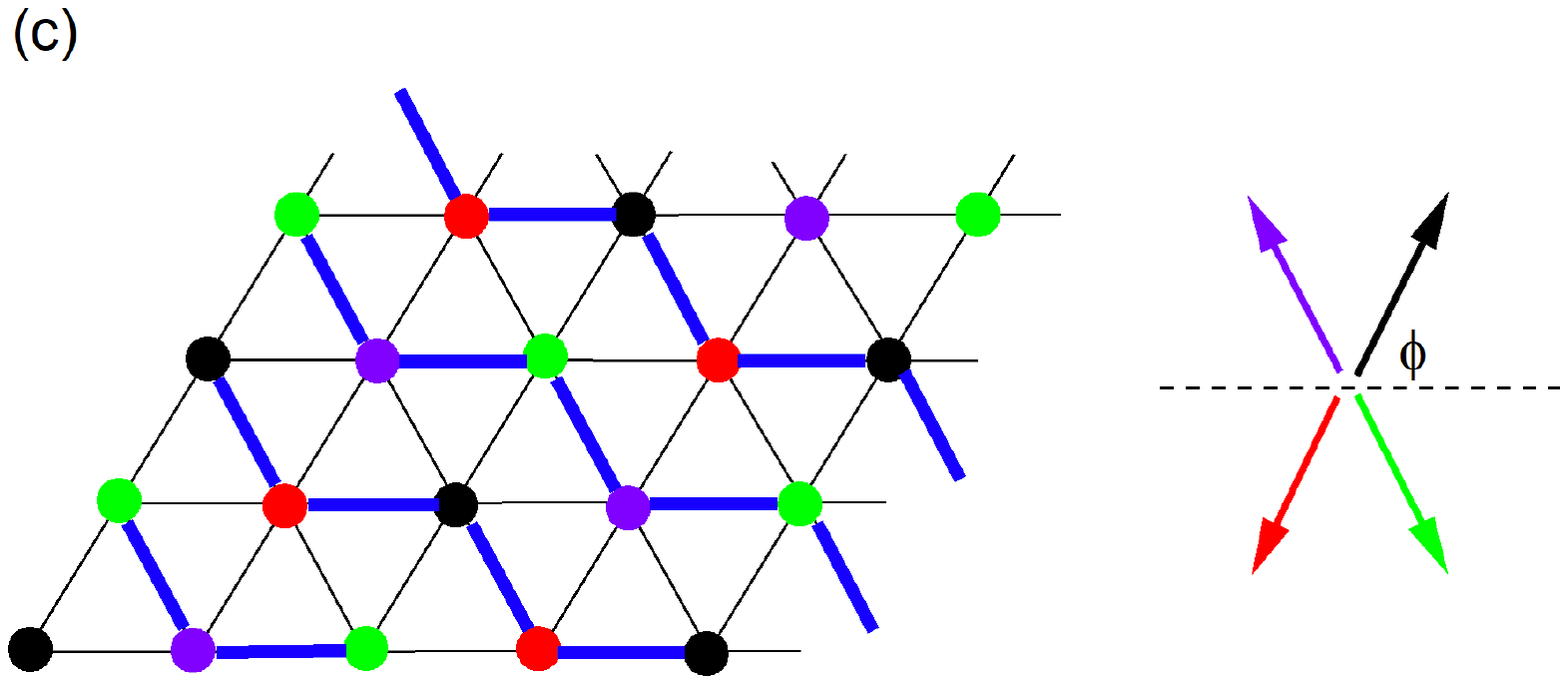}
\caption{(a)The spin structure factor of the $J_{1}-J_{4}$ model at $J_{4}=0.06$ calculated from the optimized gRVB state. (b)The spin structure factor in a classical zigzag phase. (c)Illustration of the classical zigzag phase on the triangular lattice. The thick blue bonds highlight the antiferromagnetic correlated backbones in the zigzag phase. The spin structure factor shown in (b) is calculated assuming that $\cot \phi=\frac{1}{3}$ and that the spin has a unit length.}
\end{figure}

\subsection{Comparison with the phase diagram obtained from exact diagonalization}
To provide further support to the variational results, we have performed exact diagonalization(ED) calculation on a $6\times 6$ cluster with periodic boundary condition. Here we focus on the identity representation. Using translational and point group symmetry, together with the spin rotational symmetry along the $z$-axis, we can reduce the number of symmetrized basis to 31,554,903. The Hamiltonian matrix is then diagonalized by the Lanczos method to obtain its lowest few eigenvalues. Fig.16 shows the ground state energy as a function of $J_{4}$ obtained from the ED calculation. Two features of the ground state energy curve are of particular interest to us. First, the curvature of the ground state energy is very small in both the small $J_{4}$ and the large $J_{4}$ regime, a trend in good agreement with the variational result presented above. Second, in the intermediate $J_{4}$ regime, level crossings between the ground state and the first excited state are observed, implying potential first order transition in the thermodynamic limit. Such transitions may be closely related to the appearance of the zigzag phase we obtained in the variational study. Clearly, to make more solid conclusion on this issue, more systematic ED calculation is needed.

\begin{figure}
\includegraphics[width=8.5cm]{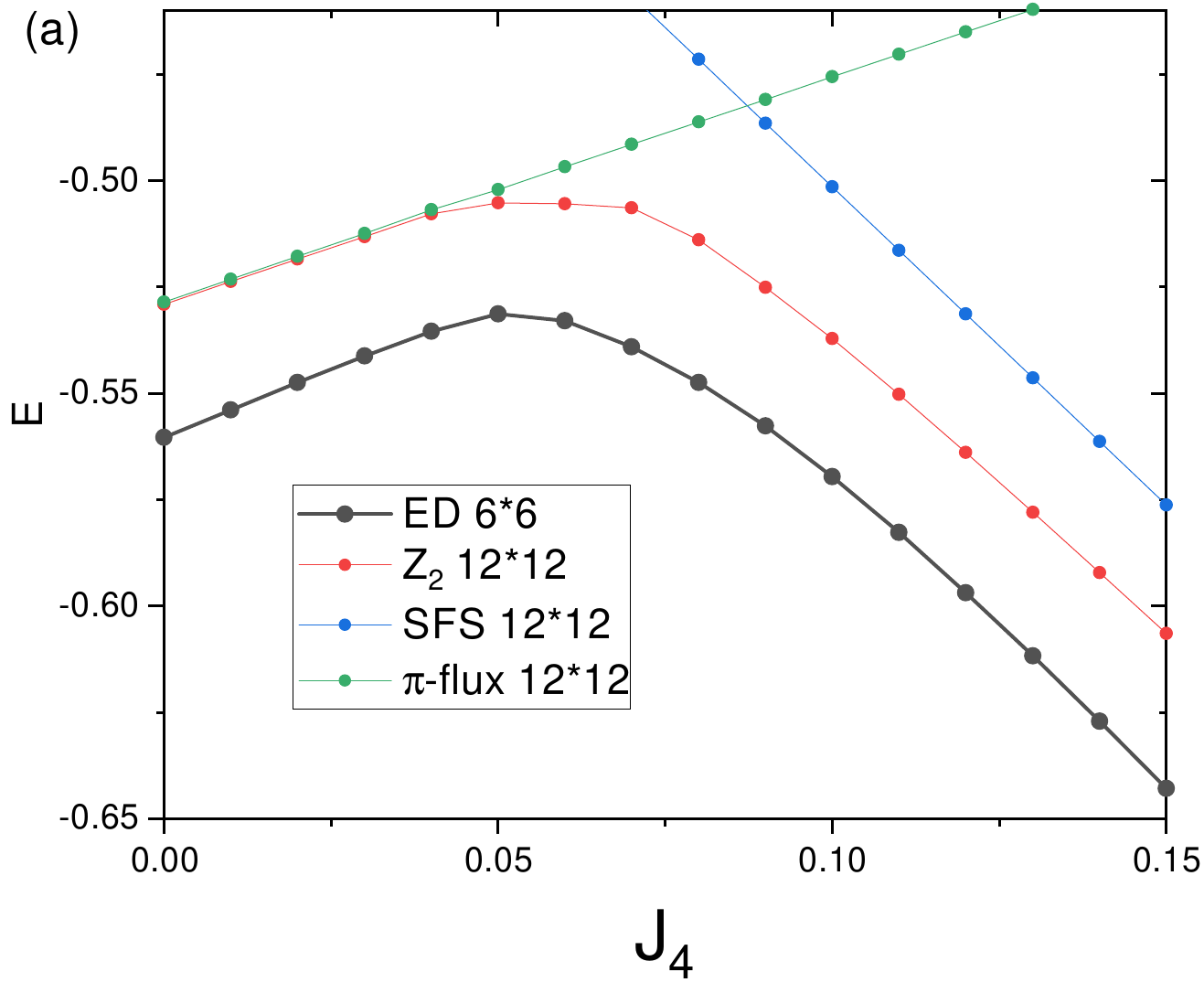}
\includegraphics[width=8.5cm]{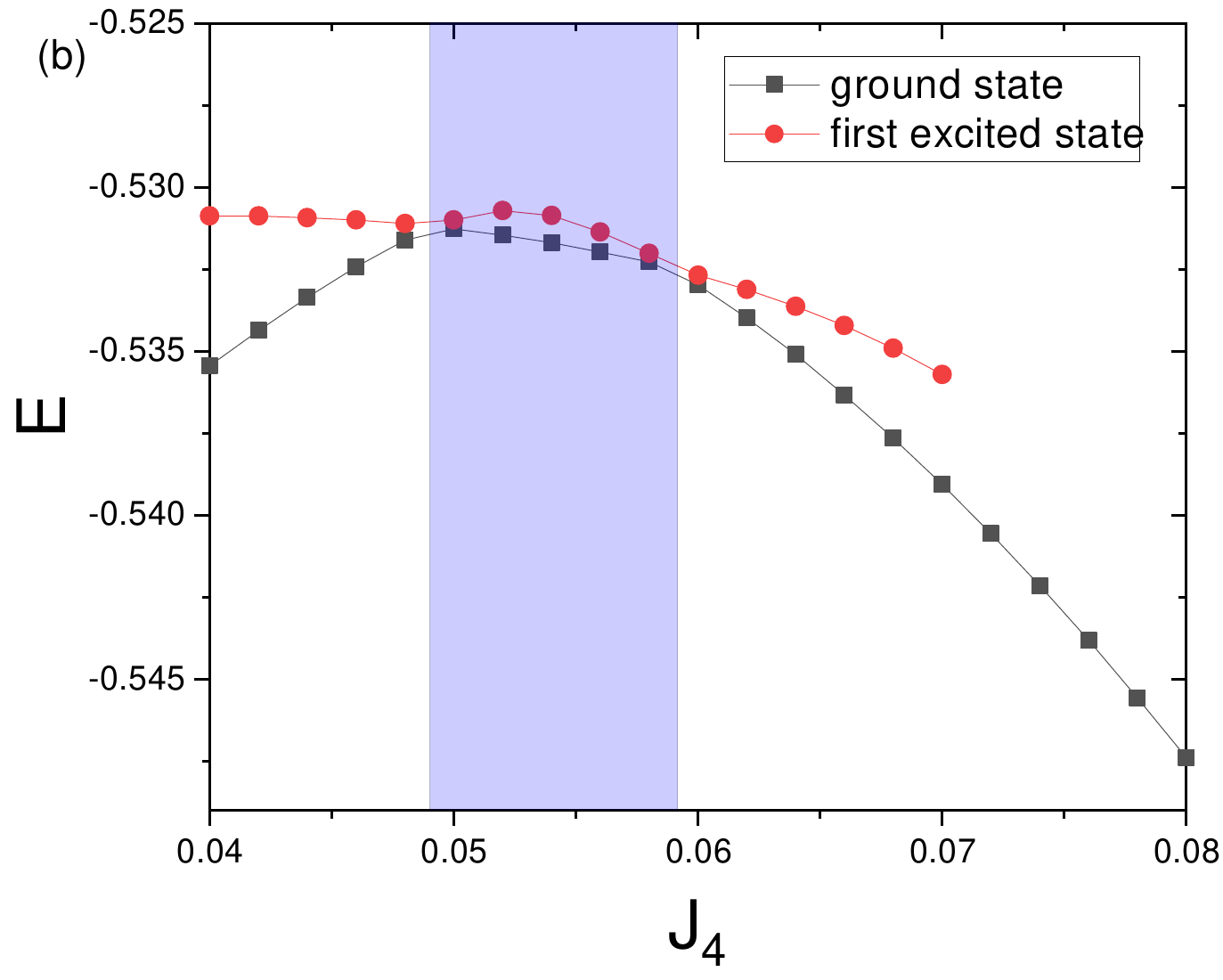}
\caption{(a)The ground state energy of the $J_{1}-J_{4}$ model calculated from exact diagonalization on a $6 \times 6$ cluster with periodic boundary condition(black dots). This is compared with the variational energies of the $\pi$-flux phase(green dots), the SFS phase(blue dots) and the optimized $Z_{2}$ RVB state(red dots) we obtained on a $12\times12$ cluster with periodic boundary condition. (b)The lowest two eigenvalues in the identity representation of the $6\times6$ cluster cross at both $J_{4}=0.049$ and $J_{4}=0.059$, implying potential first order transition there in the thermodynamic limit.}
\end{figure}

We note that the finite size effect on a $6\times6$ cluster can be very significant, especially when the considered state is gapless. For example, if we perform variational optimization on a $6\times 6$ cluster with periodic boundary condition, what we would obtain at $J_{4}=0.15$ is a symmetry broken state with a $\sqrt{3}\times\sqrt{3}$ modulation in its local spin correlation pattern( as is illustrated in Fig.17), rather the $4\times 6$ state we find on the $12\times12$ cluster. The modulation of the local spin correlation in the $\sqrt{3}\times\sqrt{3}$ state is found to be much weaker than that in the $4\times6$ state. Such a result is partially expected since the $4\times6$ state is incompatible with the periodic boundary condition of the $6\times6$ cluster.  However, when we extend the optimized ansatz of such a $\sqrt{3}\times\sqrt{3}$ state to a $12\times12$ cluster, we find that the variational energy is very bad. On the other hand, when we extend the optimized ansatz of the $4\times 6$ state on the $12\times12$ cluster to a $24\times24$ cluster, the variational energy is essentially unchanged. Fig.18 illustrates such contrasting behavior of the $\sqrt{3}\times\sqrt{3}$ state and the $4\times6$ state. Thus, the dominance of the nearly uniform $\sqrt{3}\times\sqrt{3}$ state on the $6\times6$ cluster is purely a finite size effect. This indicates that a sufficiently large cluster is needed to draw conclusion on the phase diagram of model. 

\begin{figure}
\includegraphics[width=8.5cm]{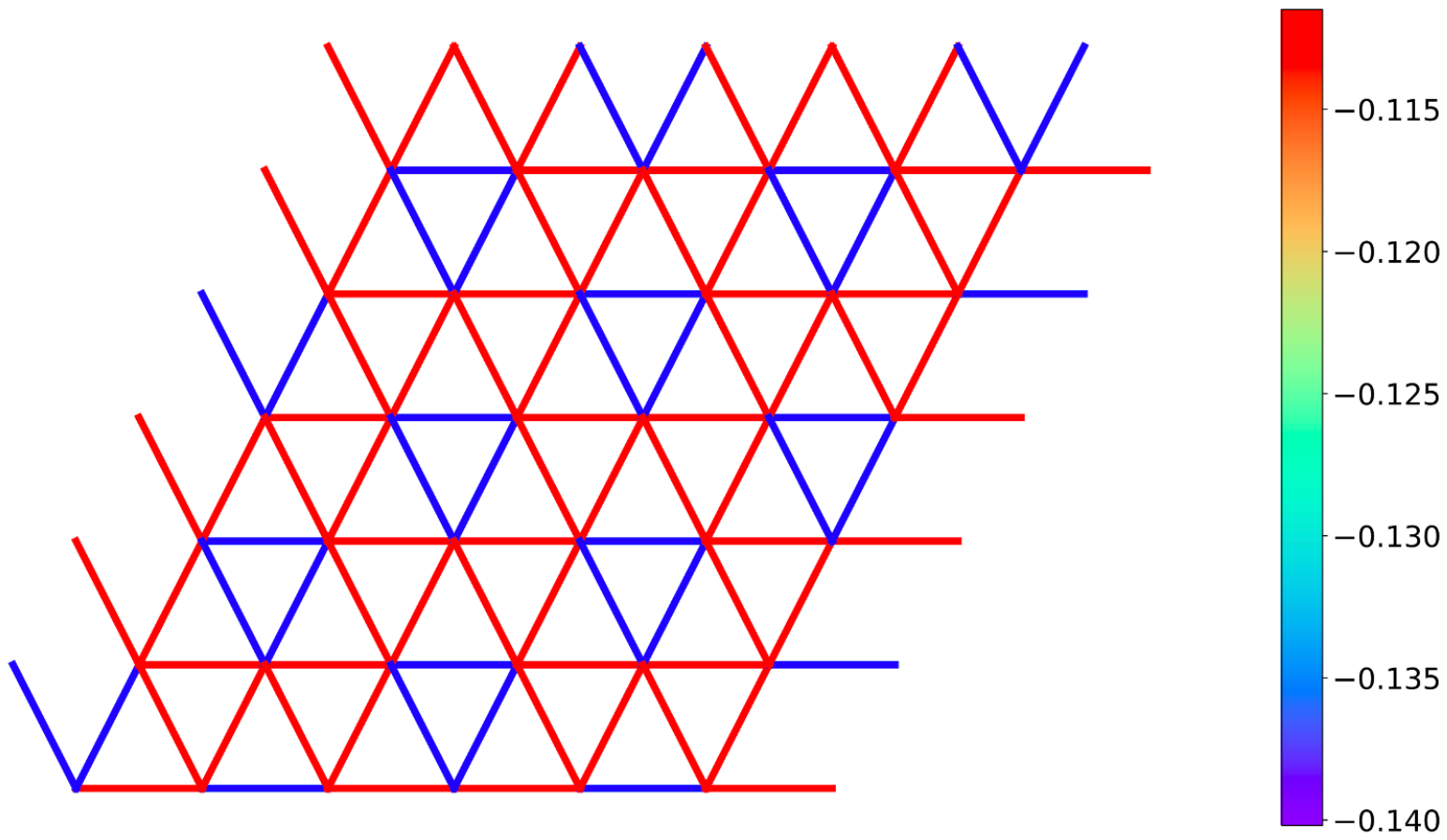}
\caption{The local spin correlation pattern of the optimized variational ground state at $J_{4}=0.15$ on a $6\times6$ cluster with periodic boundary condition. This state exhibits a $\sqrt{3}\times\sqrt{3}$ local spin correlation pattern, rather than the $4\times6$ pattern we find above on the $12\times12$ cluster. The magnitude of the modulation in the local spin correlation is also much smaller than that in the $4\times6$ state. }
\end{figure}

\begin{figure}
\includegraphics[width=8.5cm]{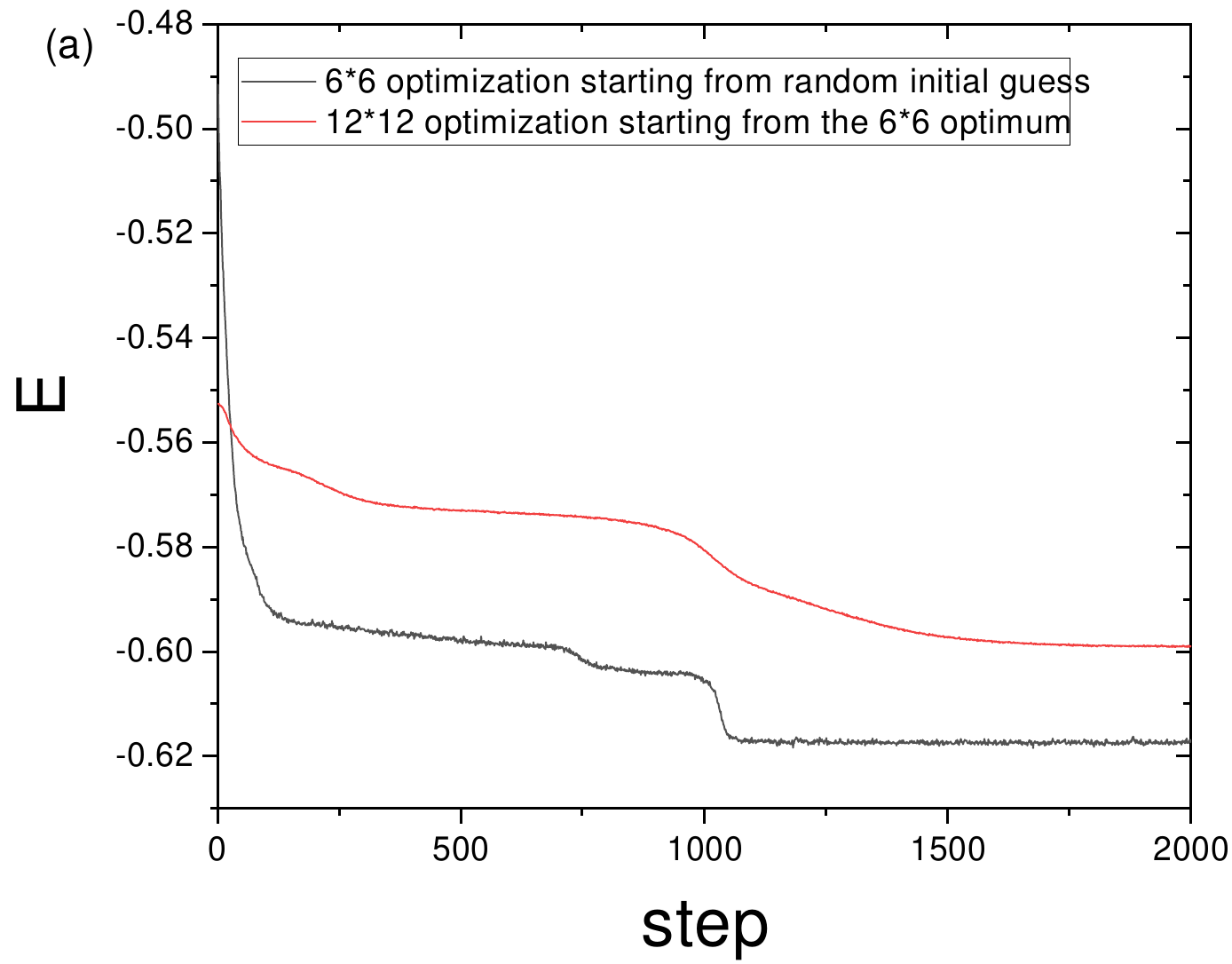}
\includegraphics[width=8.5cm]{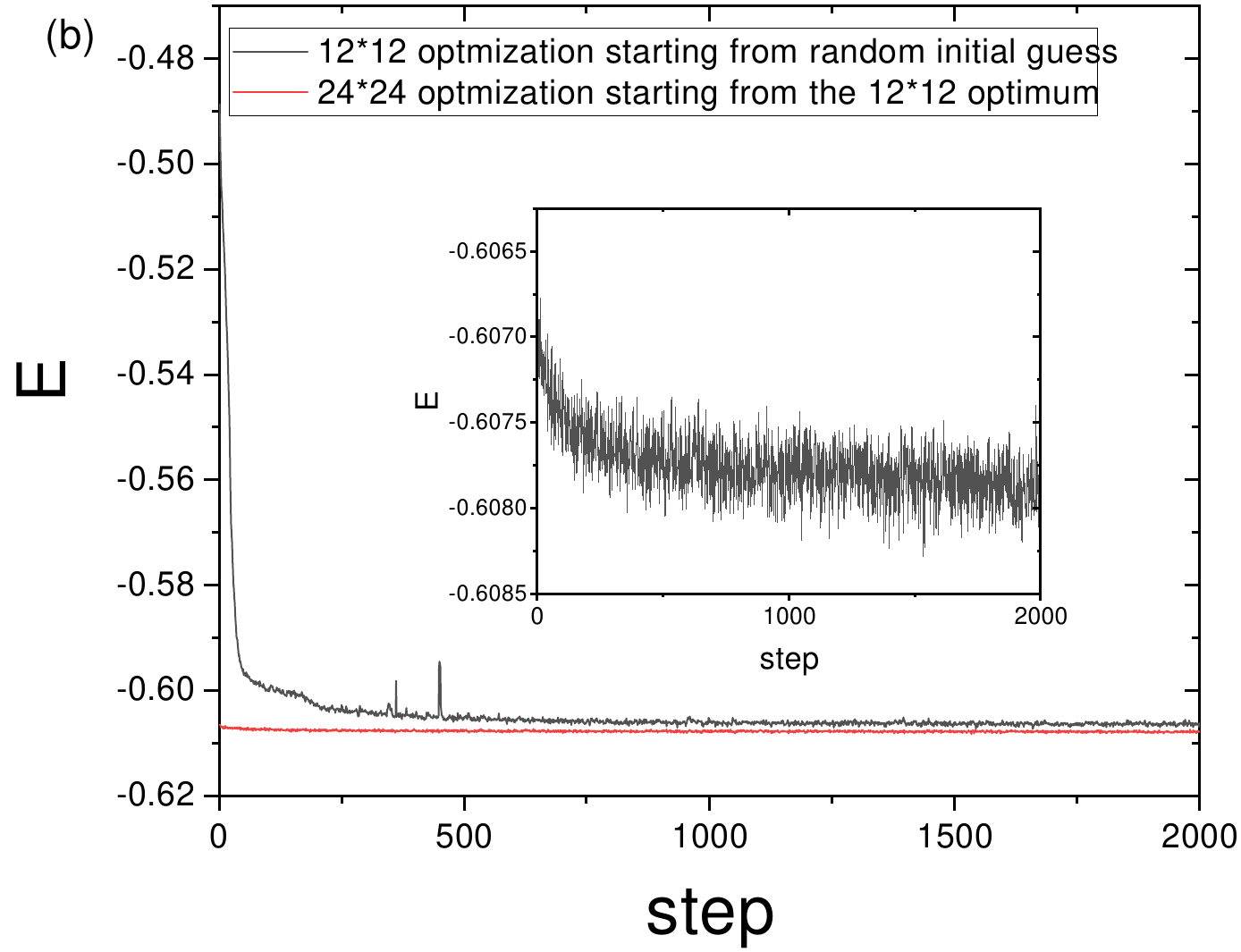}
\caption{The contrasting behavior of the $\sqrt{3}\times\sqrt{3}$ state and the $4\times6$ state during the optimization procedure at $J_{4}=0.15$. (a)While the $\sqrt{3}\times\sqrt{3}$ state is the best variational state at $J_{4}=0.15$ on a $6\times6$ cluster, it generates a very bad variational energy on a $12\times12$ cluster. (b)On the other hand, the variational energy of the $4\times6$ state is essentially unchanged when the calculation is extended from the $12\times12$ cluster to the much larger $24\times24$ cluster. This contrasting behavior indicates that the dominance of the $\sqrt{3}\times\sqrt{3}$ state on the $6\times6$ cluster is purely a finite size effect.}
\end{figure}  

\section{Discussions and conclusions}
While it is generally believed that the long-sought $U(1)$ spin liquid with a large spinon Fermi surface can be realized in the large $J_{4}$ regime of the $J_{1}-J_{4}$ model on the triangular lattice, recent DMRG simulation indicates that the expected spin liquid phase may be replaced by some symmetry breaking phase in the real phase diagram of this model. The tension between such a DMRG result and the abundant variational results on this model calls for a systematic variational study of the $J_{1}-J_{4}$ model without assuming any symmetry a prior. This is a formidable numerical task. With the increase of the number of variational parameters, the variational optimization procedure becomes increasingly tricky as a result of the abundance of local minimum in the energy landscape and/or the large fluctuation in the eigenvalues of the Hessian matrix. The $J_{1}-J_{4}$ model is a typical example in this regard. As we have shown in the last section, the SFS state is an extremely stable local minimum of the variational energy of this model as a result of its special symmetry properties.

To map out the genuine ground state phase diagram of the $J_{1}-J_{4}$ model on the triangular lattice, we have proposed several improvements on the variational optimization algorithm. The key to such improvements is a better approximation of the Hessian matrix with the gradient information. We find that the finite depth BFGS algorithm has the best balance between numerical efficiency and stability. It can often proceed further the optimization procedure when the conventional steepest descent and the stochastic reconfiguration algorithm get stuck.     

We have used the improved algorithms to optimize three kinds of Fermionic RVB state for the $J_{1}-J_{4}$ model on the triangular lattice, namely, a general RVB state whose RVB amplitudes are treated directly as variational parameters, a $U(1)$ RVB state generated from Gutzwiller projection of the ground state of a mean field Hamiltonian with only Fermion hopping terms, and a $Z_{2}$ RVB state generated from Gutzwiller projection of the ground state of a BCS-type mean field Hamiltonian. We get consistent results from all these three kinds of RVB state on the ground state phase diagram of the $J_{1}-J_{4}$ model on the triangular lattice.

From our variational optimization, we find that for $0 \le J_{4} \le 0.045$, the best variational state of the model is the well known $\pi$-flux phase on the triangular lattice. This state is the closest approximation of the 120 degree ordered phase on the triangular lattice within the subspace of the Fermionic RVB state, as can be seen from its static spin structure factor. Indeed, such a Dirac spin liquid state can be tuned continuously into a Bosonic RVB state\cite{Seiji}, which provides an extremely accurate description of the ground state of the antiferromagnetic Heisenberg model on the triangular lattice\cite{ZhangQ}. We find that the variational state for $0 \le J_{4} \le 0.045$ is almost independent of $J_{4}$, consistent with the observation that the variational energy has nearly zero curvature as a function of $J_{4}$.  Such an observation agrees well with the result obtained from ED calculation on a $6\times6$ cluster. Thus, while the $\pi$-flux phase does not posses true magnetic long range order, it nevertheless captures correctly the spin correlation pattern in the 120 degree ordered phase. We thus expect that an extended variational calculation involving magnetic long range order will not change the structure of the phase diagram qualitatively. 

For $J_{4}\ge0.09$, a regime which is thought to be the most favorable for the SFS state, we find that the optimized variational ground state exhibits a $4\times6$ modulation in its local spin correlation pattern. The magnitude of the modulation is found to be very large, ranging from nearly pure spin singlet correlation to nearly pure spin triplet correlation between nearest neighboring spins. The energy gain related to such a symmetry breaking is found to be quite large. More specifically, the energy gain of the $4\times6$ state over the SFS state is found to be about $5\%$ in the whole $J_{4}\ge 0.09$ regime. We find that this conclusion is robust on larger clusters. In addition, we find that the curvature in the variational energy is also very small in the $J_{4}\ge0.09$ regime, implying the nearly $J_{4}$-independent nature of the variational ground state. These results indicate collectively that the $4\times6$ phase is the variational ground state in the whole $J_{4}\ge0.09$ regime.   

The variational ground state of the intermediate regime of $0.045 \le J_{4}\le 0.09$ is found to exhibit zigzag modulation in its local spin correlation pattern. Depending on if an additional translational symmetry is broken or not, the intermediate regime can be further divided into a type-I zigzag phase for $0.045 \le J_{4}\le 0.055$ and a type-II zigzag phase for $0.055 \le J_{4} \le 0.09$. All these different phases are found to be connected by first order transitions. We find that the variational energy of the zigzag phase is not only much lower than that of the SFS state, but is also much lower than that of other previously proposed variational states, in particular, the nodal $d$-wave phase and the $d+id$-wave phase with quadratic band touching in spinon dispersion.

Taken all these results together, we conclude that while the SFS state is locally extremely stable, it is never the true variational ground state of the $J_{1}-J_{4}$ model on the triangular lattice. In addition, no translational symmetric spin liquid state can be stabilized in the $J_{1}-J_{4}$ model. We find that the energy advantage by breaking the translational symmetry can be very significant. Compared to such large energy gain, the tiny energy difference between the nodal $d$-wave state and the $d+id$-wave state with quadratic band touching in the spinon dispersion appears meaningless. Thus, any serious variational study of the ground state phase diagram of the $J_{1}-J_{4}$ model should take into account the possibility of translational symmetry breaking. We think that this is true not only for the particular model studied in this paper, but also applies in the variational study of general highly frustrated quantum antiferromagnetic models. The BFGS algorithm proposed in this paper thus has much broader range of applications in future study of quantum spin liquids. 

Another lesson that we can learn from our result is that the finite size effect may significantly distort the phase diagram of a highly frustrated spin model on a finite cluster, since the model may develop symmetry breaking pattern with a rather large unit cell in its ground state. To extract the genuine behavior of the model in the thermodynamic limit, sufficiently large system(or sufficiently wide system in the case of DMRG simulation) is needed. This calls for new developments in the variational optimization algorithm and the DMRG algorithm.

Finally, while the SFS state is unstable against the $4\times6$ state in the large $J_{4}$ regime, it may still be stabilized in real materials by additional couplings that frustrate the $4\times6$ modulation pattern. A thorough study involving other major spin couplings that can be generated from the strong coupling expansion of the Hubbard model on the triangular lattice\cite{Schmidt} is necessary to fully address this problem.  We leave such a study to future works.

\begin{acknowledgments}
We acknowledge the support from the National Natural Science Foundation of China(Grant No.12274457). 
\end{acknowledgments}


\begin{thebibliography}{}
\bibitem{note1}We note that the spin correlation function in the zigzag phase is not translational invariant. Thus its Fourier transform is not diagonal in wave vector. The spin structure factor we plot in Fig.15 corresponds to the diagonal part of the Fourier component of the spin correlation function.
\end{thebibliography}
\end{document}